\documentclass[twocolumn,showpacs,aps,prd]{revtex4}

\usepackage{graphicx}
\usepackage{dcolumn}
\usepackage{amsmath}
\usepackage{epsfig}
\usepackage{verbatim}
\usepackage{mathrsfs}

\input babarsym

\newcommand{\BABARPubYear}    {10}
\newcommand{\BABARPubNumber}  {018}

\newcommand{\SLACPubNumber} {14308}

\def\figurebox#1#2#3{%
    \def\arg{#3}%
    \ifx\arg\empty
    {\hfill\vbox{\hsize#2\hrule\hbox to #2{\vrule\hfill\vbox to #1{\hsize#2\vfill}\vrule}\hrule}\hfill}%
    \else
    {\hfill\epsfbox{#3}\hfill}%
    \fi}

\def\babar{\mbox{\slshape B\kern-0.1em{\smaller A}\kern-0.1em
    B\kern-0.1em{\smaller A\kern-0.2em R}}}
\def\DDK {\ensuremath{\Dbar^{(*)} D^{(*)}  K}}
\def\DDKpm {\ensuremath{\Dbar^{(*)} D^{(*)} \Kpm}}

\def\DsOneRes {\ensuremath{D^+_{s1}(2536)}}
\def\DsJRes {\ensuremath{D_{sJ}(2700)}}
\def\PsiRes {\ensuremath{\psi(3770)}}
\def\XRes {\ensuremath{X(3872)}}

\newcommand{\BToDDK}{\ensuremath{B\to \Dbar^{(*)} D^{(*)} K}}
\def\W      {\ensuremath{W}\xspace}
\def\to {\rightarrow}

\def\modei{\ensuremath{\Bz\to \Dm \Dz \Kp}}
\def\modeii{\ensuremath{\Bz\to \Dzb \Dz K^0}}
\def\modeiii{\ensuremath{\Bz\to \Dm \Dp K^0}}
\def\modeiv{\ensuremath{\Bz\to \Dm \Dstarz \Kp}}
\def\modev{\ensuremath{\Bz\to \Dstarm \Dz \Kp}}
\def\modevi{\ensuremath{\Bz\to \Dzb \Dstarz K^0 + \Dstarzb \Dz K^0}}
\def\modevii{\ensuremath{\Bz\to \Dm \Dstarp K^0 + \Dstarm \Dp K^0}}

\def\modeviii{\ensuremath{\Bz\to \Dstarm \Dstarz \Kp}}
\def\modeix{\ensuremath{\Bz\to \Dstarzb \Dstarz K^0}}
\def\modex{\ensuremath{\Bz\to \Dstarm \Dstarp K^0}}
\def\modexi{\ensuremath{\Bu\to \Dzb \Dz \Kp}}
\def\modexii{\ensuremath{\Bu\to \Dzb \Dp K^0}}
\def\modexiii{\ensuremath{\Bu\to \Dm \Dp \Kp}}
\def\modexiv{\ensuremath{\Bu\to \Dstarzb \Dz \Kp}}
\def\modexv{\ensuremath{\Bu\to \Dstarzb \Dp K^0}}
\def\modexvi{\ensuremath{\Bu\to \Dstarm \Dp \Kp}}
\def\modexvii{\ensuremath{\Bu\to \Dzb \Dstarz \Kp}}
\def\modexviii{\ensuremath{\Bu\to \Dzb \Dstarp K^0}}
\def\modexix{\ensuremath{\Bu\to \Dm \Dstarp \Kp}}
\def\modexx{\ensuremath{\Bu\to \Dstarzb \Dstarz \Kp}}
\def\modexxi{\ensuremath{\Bu\to \Dstarzb \Dstarp K^0}}
\def\modexxii{\ensuremath{\Bu\to \Dstarm \Dstarp \Kp}}

\def\mes    {\mbox{$m_{\rm ES}$}\xspace}
\def\de    {\mbox{$\Delta E$}\xspace}

\newcommand\CellTop{\rule{0pt}{2.35ex}}

\newcommand\CellTopThree{\rule{0pt}{2.8ex}}
\newcommand\CellTopFour{\rule{0pt}{3.6ex}}

\begin{document}

\preprint{\babar-PUB-\BABARPubYear/\BABARPubNumber}
\preprint{SLAC-PUB-\SLACPubNumber}

\begin{flushleft}
\babar-PUB-\BABARPubYear/\BABARPubNumber\\
SLAC-PUB-\SLACPubNumber\\
\end{flushleft}

\title{
{\large \bf Measurement of the \boldmath{$B \to \DDK$} branching fractions } }

%
\author{P.~del~Amo~Sanchez}
\author{J.~P.~Lees}
\author{V.~Poireau}
\author{E.~Prencipe}
\author{V.~Tisserand}
\affiliation{Laboratoire d'Annecy-le-Vieux de Physique des Particules (LAPP), Universit\'e de Savoie, CNRS/IN2P3,  F-74941 Annecy-Le-Vieux, France}
\author{J.~Garra~Tico}
\author{E.~Grauges}
\affiliation{Universitat de Barcelona, Facultat de Fisica, Departament ECM, E-08028 Barcelona, Spain }
\author{M.~Martinelli$^{ab}$}
\author{A.~Palano$^{ab}$ }
\author{M.~Pappagallo$^{ab}$ }
\affiliation{INFN Sezione di Bari$^{a}$; Dipartimento di Fisica, Universit\`a di Bari$^{b}$, I-70126 Bari, Italy }
% ...
\collaboration{The \babar\ Collaboration}
\noaffiliation

\date{\today}

\begin{abstract}
\noindent
We present a measurement of the branching fractions of the 22 decay channels of the $B^0$ and
$B^{+}$ mesons to \DDK, where the $D^{(*)}$ and $\Dbar^{(*)}$ mesons are fully
reconstructed.
Summing the 10 neutral modes and the 12 charged modes, the branching fractions are found to be
 $\BR(B^0 \to \DDK) = (3.68 \pm 0.10 \pm 0.24)\%$ and $\BR(B^+ \to \DDK) = (4.05 \pm 0.11 \pm 0.28)\%$, where the first uncertainties are statistical and the second systematic.
The results are based on $429~\mathrm{fb}^{-1}$ of data containing $471\times10^6\, B\Bbar$ pairs collected at the $\FourS$ resonance with the \babar\ detector at the SLAC National Accelerator Laboratory.
\end{abstract}


\maketitle


\section{Introduction}
\label{sec:intro}

\noindent
In this article, we report on the measurement of the branching fractions of the 22 decays of charged and neutral \B mesons to \DDK\ final states (Table~\ref{DDKModes}): $D^{(*)}$ is either a \Dz, \Dstarz, \Dp or \Dstarp, $\Dbar^{(*)}$ is the charge conjugate of $D^{(*)}$ and $K$ is
either a \Kp or a $K^0$. Both $\Dbar^{(*)}$ and $D^{(*)}$ are fully reconstructed. Charge
conjugate reactions are assumed throughout this article.

In the past, the values measured for hadronic decays of the $B$ meson were in disagreement with the expectations based on the $B$ semileptonic branching fraction due to the inconsistency originating from the number of charmed hadrons per $B$ decay (charm counting)~\cite{bigibrowder}. The $\b \to \c \cbar \s$ transition in $B$ decays was believed to be dominated by $B \to D_s \, X $, $B \to (\c\cbar)\, X $, and $B \to \Xi_c \, X $ final states, where $X$ represents any particles.  However, it was realized~\cite{buchalla} that an enhancement in the $\b \to \c \cbar \s$ transition was needed to resolve the theoretical discrepancy with the $B$ semileptonic branching fraction. Buchalla \emph{et al.}~\cite{buchalla} predicted sizeable branching fractions for decays of the form $B \to \Dbar^{(*)} D^{(*)} K\,(X)$.
Experimental evidence in support of this picture soon appeared in the literature~\cite{cleoaleph}, including a study by \babar\ using 76~\invfb\ of data where the Collaboration reported the observations or the limits on the 22 decays \BToDDK~\cite{ref:patrick}. The aggregate branching fraction measurements were $\BR(B^0 \to \DDK) = (4.3 \pm 0.3 \pm 0.6) \%$ and $\BR(B^+ \to \DDK) =
(3.5 \pm 0.3 \pm 0.5) \%$, where the first uncertainties are statistical and the second
systematic. This result may be compared with the wrong-sign $D$ production ($\b \to \c \cbar \s$ transition containing a \Db\ meson) that \babar\ studied using inclusive $B$ decays to final states containing at least one charm particle~\cite{fabrice}. The wrong-sign $D$ production was found to be $\BR(\Bzb \to \Dbar X) = (10.4 \pm 1.9)\%$ and
$\BR(\Bm \to \Dbar X) = (11.1 \pm 0.9)\%$.
In addition, \babar\ found a value of the total charm yield per $B$ decay consistent with the one derived from the semileptonic branching fraction, which solved the longstanding problem of the charm counting.

Furthermore, \DDK\ events are interesting for a variety of studies. These events can be used to investigate isospin relations and to extract a measurement of the ratio of $\Y4S \to B^+B^-$
and $\Y4S \to B^0\Bzb$ decays~\cite{ref:marco}. It was shown theoretically that the time-dependent rate for $\Bz\to D^{(*)-} D^{(*)+} \KS$ decays can be used to measure $\sin 2\beta$ and $\cos 2\beta$~\cite{datta}.
\babar\ used the mode $\Bz\to \Dstarm \Dstarp \KS$ with 209~\invfb\ of data
to perform a time-dependent \CP asymmetry measurement to determine the sign of $\cos2\beta$, under some theoretical and resonant structure assumptions~\cite{chunhui}.
The Belle Collaboration also published a similar analysis~\cite{ref:bellemodex}.
Although the resonant states are not studied in our paper, it is worth recalling that many $D^{(*)}K$ and $\Db^{(*)}D^{(*)}$ resonant processes are at play in the studied decay channels. Using $B \to \DDK$ final states, \babar\ and Belle observed and measured the properties of the resonances \DsOneRes, \DsJRes, \PsiRes, and \XRes~\cite{ref:belleDsJ, ref:myPaper, ref:tagir}.

\begin{figure*}[htb]
\begin{center}
\epsfig{file=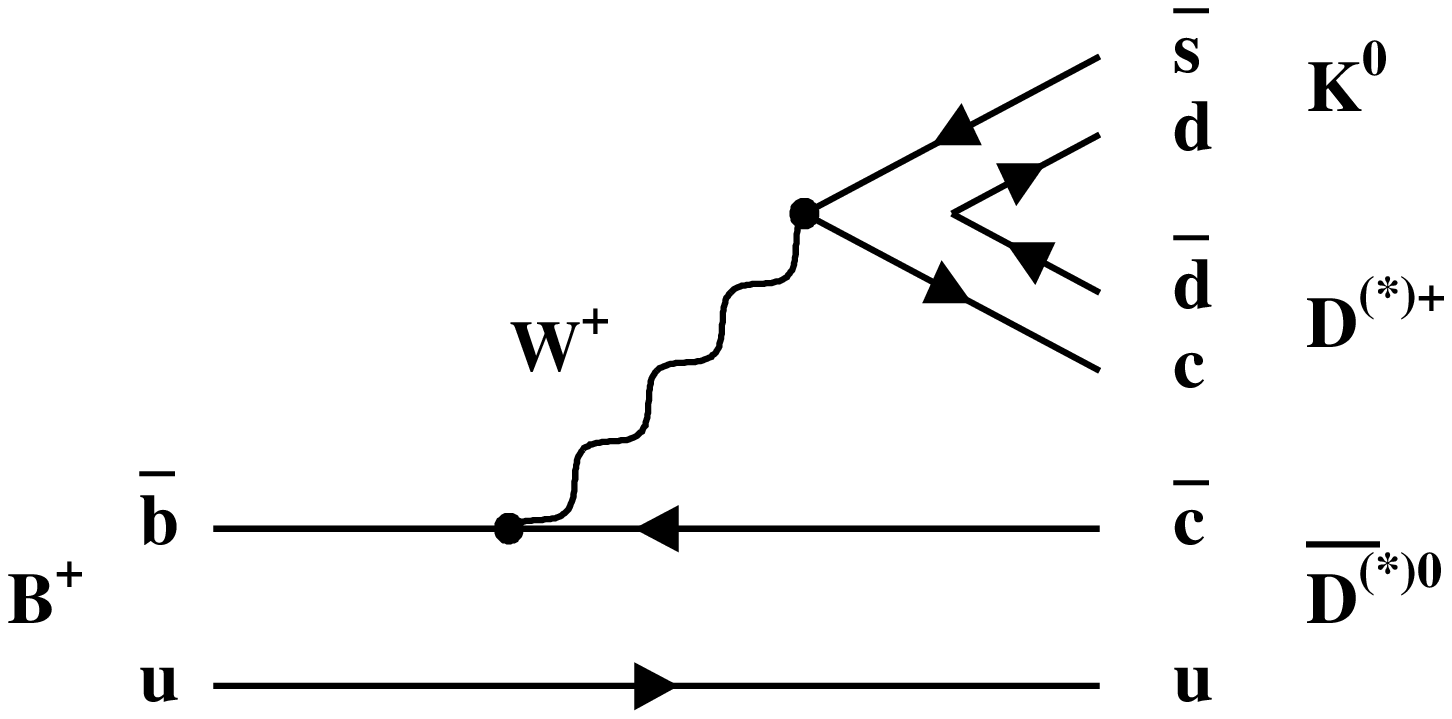,width=5.8cm}
\epsfig{file=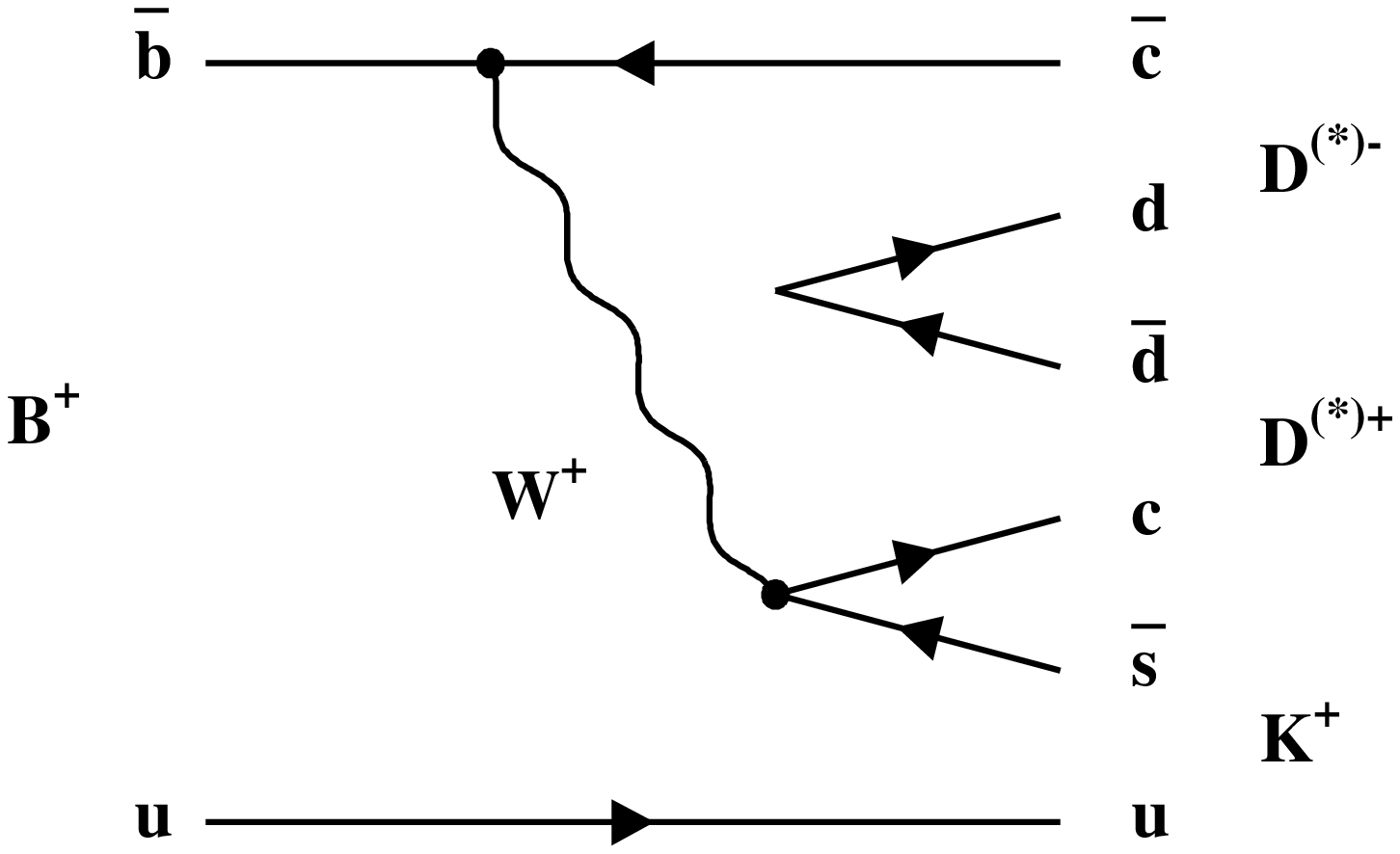,width=5.8cm}
\epsfig{file=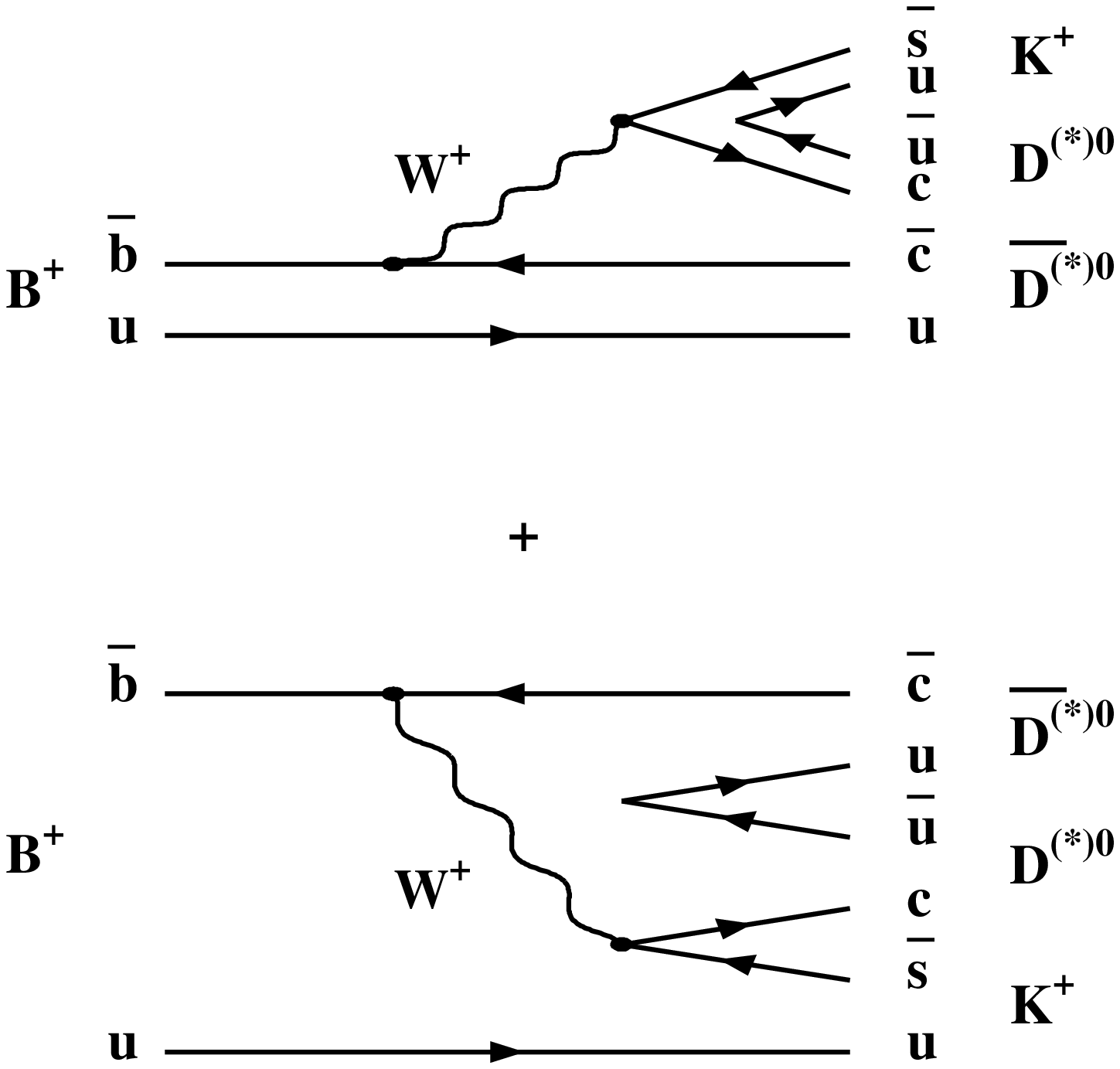,width=5.8cm}
\epsfig{file=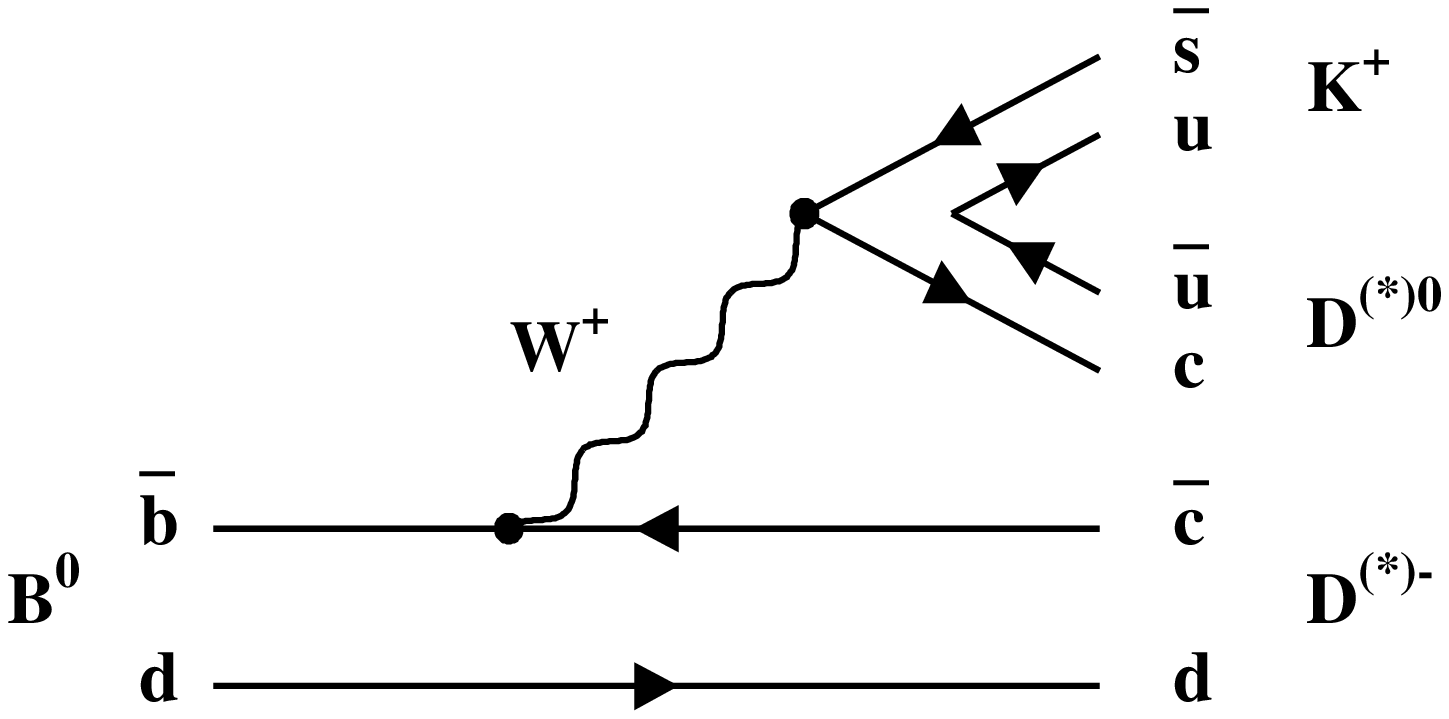,width=5.8cm}
\epsfig{file=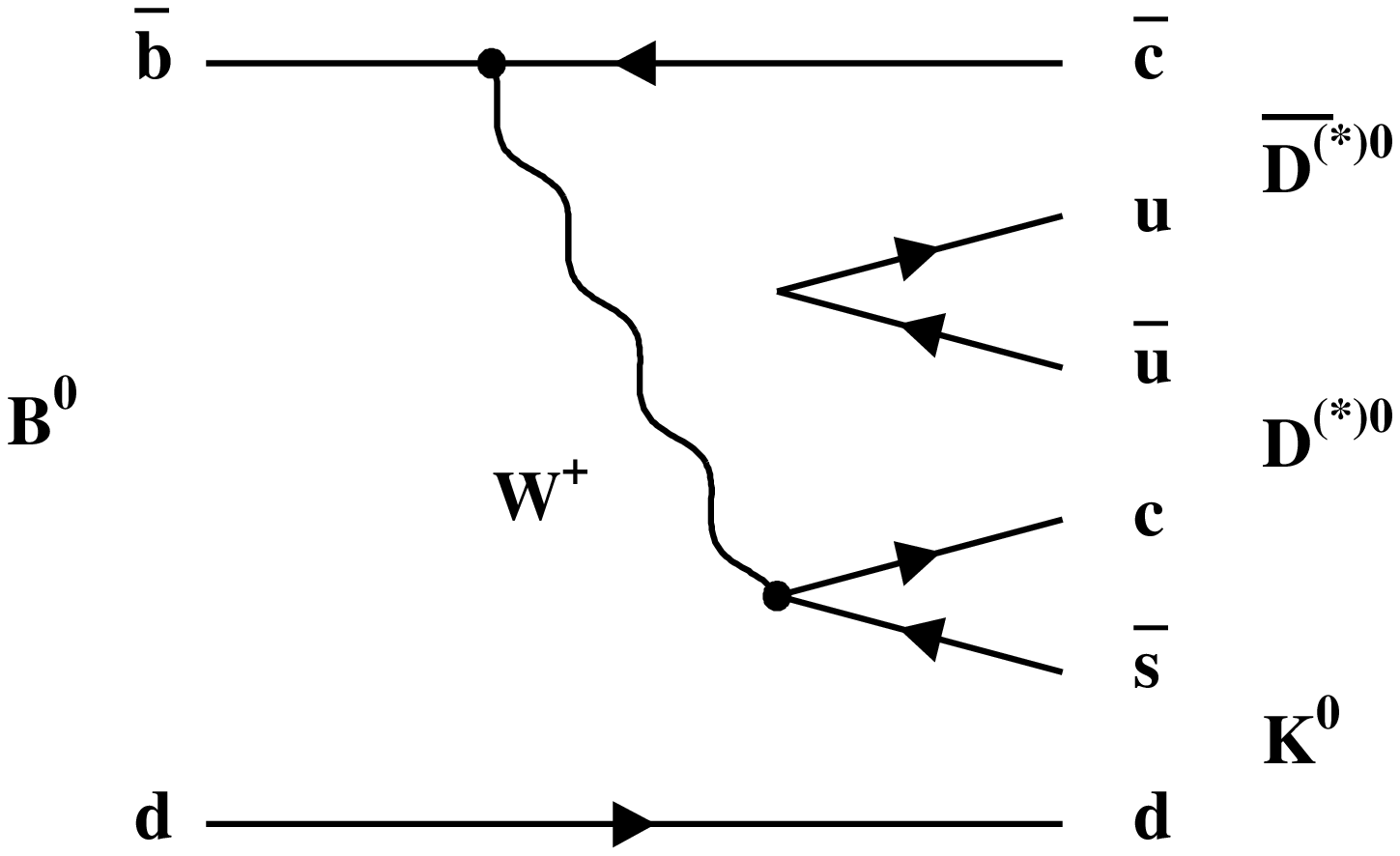,width=5.8cm}
\epsfig{file=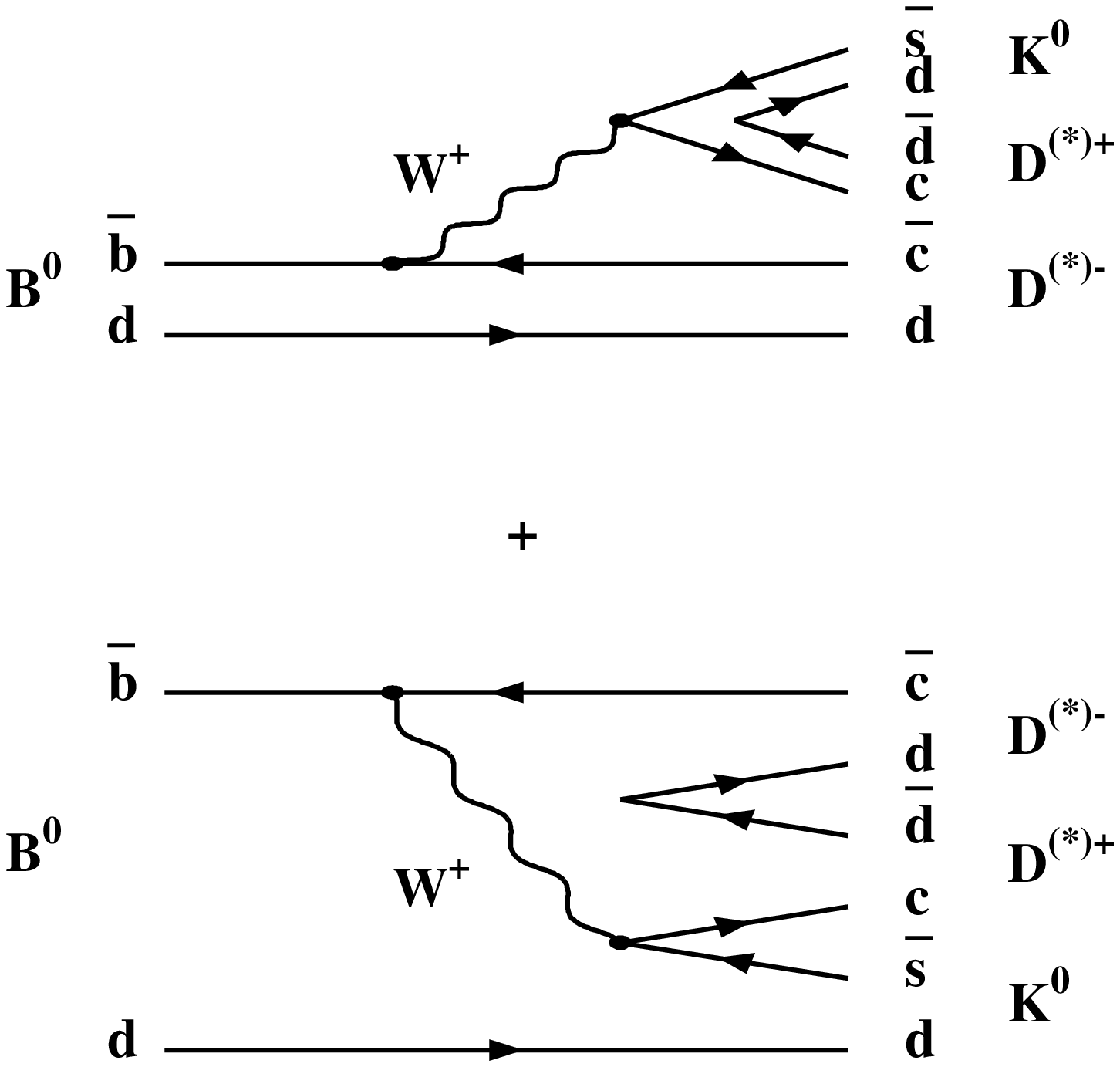,width=5.8cm}
\caption{Top left: external W-emission amplitude for the decays $B^+ \to \Dbar^{(*)0}
D^{(*)+} K^0$. Top center: internal W-emission
amplitude for the decays $B^+ \to D^{(*)-} D^{(*)+} K^+$. Top right: external+internal W-emission amplitudes for the decays $B^+ \to \Dbar^{(*)0} D^{(*)0} K^+$. Bottom row: same as top row respectively for $B^0 \to D^{(*)-} D^{(*)0} K^+$, $B^0 \to \Dbar^{(*)0} D^{(*)0} K^0$ and $B^0 \to D^{(*)-} D^{(*)+} K^0$.}
 \label{fig:diagrams}
\end{center}
\end{figure*}

The \BToDDK\ decays can proceed through external \W-emission
and internal \W-emission amplitudes, also called color-suppressed amplitudes.
As Fig.~\ref{fig:diagrams} illustrates, some decay modes proceed through only one of these amplitudes while others proceed through both.

In this paper, we update with the full \babar\ data sample our previous measurement~\cite{ref:patrick} of the branching fractions for
the 22 $B\to \Dbar^{(*)} D^{(*)} K^0$ and $B\to \Dbar^{(*)} D^{(*)} \Kp$ decays.
We benefit from several improvements with respect to this previous measurement:
\begin{itemize}
\item the integrated luminosity used for this analysis is more than 5 times larger,
\item the track reconstruction and particle identification algorithms have been improved (in purity and efficiency),
\item the efficiency of the selection of signal events has been increased,
\item the fit uses a more accurate signal parametrization,
\item the peaking background is taken into account in the fit,
\item we use a method that is insensitive to the possible resonant structure in the final states.
\end{itemize}
Measuring the 22 modes altogether allows to avoid biases in the branching fraction measurement by correctly taking into account the cross-feed events, which are events from one mode being reconstructed as a candidate for another mode.

\begin{table}[htb]
 \begin{center}
 \caption{The 22 $B\to \Dbar^{(*)} D^{(*)} K$ decay modes. The modes $\Bz \to \Dm \Dstarp K^0$
and $\Bz \to \Dstarm \Dp K^0$ are combined together since they are not experimentally
distinguishable. The same applies to the modes $\Bz\to \Dzb \Dstarz K^0$ and
$\Bz \to \Dstarzb \Dz K^0$ which are also combined together.}
 \label{DDKModes} \vskip 0.2cm
 \begin{tabular*}{0.48\textwidth}{ll}
 \hline
 \hline
\CellTop Neutral $B$ mode & Charged $B$ mode \\
 \hline
\CellTop
\modei & \modexii \\
\modeiv & \modexviii \\
\modev & \modexv \\
\modeviii & \modexxi \\
\modeiii & \modexi \\
\modevii~~~~~~ & \modexvii \\
         & \modexiv \\
\modex & \modexx \\
\modeii & \modexiii \\
\modevi & \modexix \\
        & \modexvi \\
\modeix  & \modexxii \\
\hline
\hline
 \end{tabular*}
 \end{center}
\end{table}

\section{The \babar\ detector and data sample}

\noindent
The data were recorded by the \babar\ detector at the PEP-II asymmetric-energy $e^+e^-$ storage ring operating at the SLAC National Accelerator Laboratory. We analyze the complete \babar\ data sample collected at the $\FourS$ resonance corresponding to an integrated luminosity of 429~\invfb, giving $N_{\BB} = (470.9 \pm 0.1 \pm 2.8)
\times 10^6$ \BB pairs produced, where the first uncertainty is statistical and the second
systematic.

The \babar\ detector is described in detail elsewhere~\cite{babar}. Charged particles are detected and their momenta measured with a five-layer silicon vertex tracker and a 40-layer drift chamber in a 1.5 T axial magnetic field. Charged particle identification is based on the measurements of the energy loss in the tracking devices and of the Cherenkov radiation in the ring-imaging detector. The energies and locations of showers associated with photons are measured in
the electromagnetic calorimeter. Muons are identified by the instrumented magnetic-flux return, which is located outside the magnet.

We employ a Monte Carlo (MC) simulation to study the relevant backgrounds and estimate the selection efficiencies.
We use \textsc{EVTGEN}~\cite{ref:evtgen} to model the kinematics of $B$ mesons and \textsc{JETSET}~\cite{ref:jetset} to model continuum processes, $\epem\to\qqbar$ ($q=u,d,s,c$). The \babar\ detector and its response to particle interactions are modeled using the \textsc{GEANT4}~\cite{ref:geant4} simulation package.

\section{\boldmath ${B}$ candidate selection}

\noindent
We reconstruct the $B^0$ and $B^+$ mesons in the 22 \DDK\ modes.
The level of background widely varies among the signal channels, even within a
specific $B$ mode depending on the $D$ meson decay type. A different optimization of the selection
criteria is implemented for each of the final states. The optimization determines the selection which maximizes $S/\sqrt{S+B}$, where $S$ and $B$ are the expected number of events for the signal and for the background
in the signal region, based respectively on signal and background MC simulated events. The branching fractions for the computation of $S$ are taken from our previous measurements of these channels~\cite{ref:patrick}.

We identify charged kaons using either loose or tight criteria depending on the decay mode. The loose criterion is typically 98\% efficient with pion misidentification rates at the 15\% level, while the tight criterion is 85\% efficient with a misidentification around 2\%. We use only the \KS\ meson when a neutral $K$ meson is present in the final state.
The \KS candidates are reconstructed from two oppositely charged tracks assumed to be pions consistent with coming from a common vertex and having
an invariant mass within $\pm 9.5\mevcc$ of the nominal \KS mass~\cite{ref:pdg}. The displacement of the \KS vertex in the plane transverse to the beam axis is required to be at least 0.2\cm.

The \piz candidates are reconstructed from pairs of photons with energies
$E_\gamma>30\mev$ in the laboratory frame that have an invariant mass of
$115<m_{\gaga}<150\mevcc$.

We reconstruct $D$ mesons in the modes $\Dz\to \Km\pip$, $\Km\pip\piz$,
$\Km\pip\pim\pip$, and $\Dp\to \Km\pip\pip$. The $K$ and $\pi$ tracks are required to
originate from a common vertex. The invariant masses of the $D$ candidates are required
to lie within $\pm 2.5\sigma_D$ of the measured $D$ mass, where $\sigma_D$ is the $D$ invariant mass resolution.
This resolution is
measured to be 5.8 \mevcc\ for $D^0 \to K^- \pi^+$, 9.5 \mevcc\ for $D^0 \to K^- \pi^+
\pi^0$, 4.7 \mevcc\ for $D^0 \to K^- \pi^+ \pi^- \pi^+$, and 4.2 \mevcc for $D^+ \to K^-
\pi^+ \pi^+$. To reduce the combinatorial background, for some of the $B$ decays involving $D^0 \to K^- \pi^+ \pi^0$, we use
the distribution of events in the Dalitz plot of the squared invariant masses $m^2(K^- \pi^+) \times m^2(K^- \pi^0)$ where we select events that are located in the enhanced regions dominated by the $K^*(892)^+$, $K^*(892)^0,$ and $\rho(770)^+$ resonances~\cite{ref:kpipi0}.

The $D^*$ candidates are reconstructed in the decay modes $D^{*+}\to \Dz\pip$, $D^{*+}\to
D^+\piz$, $\Dstarz\to \Dz\piz$, and $\Dstarz\to \Dz\g$. The \piz and the \pip candidates must have
a momentum smaller than 450\mevc in the \FourS rest frame, while the \g energy in the laboratory
frame must be larger than 100\mev. The mass difference between the \Dstarp and $D$
candidates is required to be within $\pm 3 \mevcc$ of the nominal value~\cite{ref:pdg}. For
$D^{*0}$ meson decays, the mass difference between the \Dstarz and $D^0$ candidates is required to lie between 138 and 146 \mevcc for \Dstarz $\to$ \Dz \piz and between 130 and 150 \mevcc for \Dstarz $\to$
\Dz $\gamma$.

The $B$ candidates are reconstructed by combining a $\Dbar^{(*)}$, a $D^{(*)}$ and a $K$
candidate in one of the 22 modes. For modes involving two \Dz mesons, at least one of
them is required to decay to $\Km \pip$, except for the decay modes $D^{*-}D^{*+}K^0$,
$D^{*-}D^{*+}\Kp$, and $D^{*-}\Dz\Kp$, which have lower background and for which all
combinations are accepted. For modes containing a $D^{*+}$ meson, we look only to the decay
$D^{*+} \to D^0 \pi^+$, except for the modes containing $D^{*-}D^{*+}$, where we also
reconstruct $D^{*+} \to D^+ \pi^0$. A mass-constrained kinematic fit is
applied to the intermediate particles (\Dstarz, $D^{*+}$, \Dz, \Dp, \KS, \piz) to improve
their momentum resolution.

To suppress the continuum background, we remove
events with $R_2>0.3$ (where $R_2$ is the ratio of the second to zeroth Fox-Wolfram
moments of the event~\cite{fox}) and events with $|\cos (\theta_B)| > 0.9$ (where
$\theta_B$ is the angle between the thrust axis of the candidate decay and the thrust axis
of the rest of the event).

Two kinematic variables are used to isolate the $B$-meson signal.
The first variable is the beam-energy-substituted mass defined as
\begin{equation}
\mes = \sqrt{{\left( {\frac {{s / 2}
+\vec{p}_0.\vec{p}_B} {E_0}}  \right)^2}-\vert \vec{p}_B \vert
^2},
\end{equation}
where $\sqrt{s}$ is the $e^+e^-$ center-of-mass energy. For the momenta
$\vec{p}_0$, $\vec{p}_B$  and the energy $E_0$, the subscripts $0$ and
$B$ refer to the $e^+e^-$ system and the reconstructed $B$ meson,
respectively. The other variable is $\de$, the difference between the reconstructed
energy of the $B$ candidate and the beam energy in the $e^+e^-$
center-of-mass frame. Signal events have \mes\ compatible with the known $B$-meson
mass~\cite{ref:pdg} and \de\ compatible with 0~\mev, within their
respective resolutions. At this stage, we keep only events which satisfy $\mes>5.20 \gevcc$.

We obtain a few signal $B$ candidates per event
on average. When the final state contains no $D^*$ meson, we get 1.0 to 1.3 candidates
per event depending on the specific mode, 1.3 to 1.9 candidates per event for final
states containing one $D^*$ meson, and 1.7 to 2.1 candidates per event when the final
state contains two $D^*$ mesons (except for \modexx\ with 2.9 candidates per event). If
more than one candidate is selected in an event, we retain the one with the smallest value of $|\de|$ (``best candidate selection'').
According to MC studies, this criterion finds the correct candidate when this one is present in the candidate list in more than 95\% of the cases for final states with no $D^{*0}$ meson and more than 80\% of the cases for modes with one or two neutral $D^{*}$ mesons.
We keep only events with $|\de|<E_c$ with $E_c$ varying from 7 \mev to 56 \mev depending on the decay mode of the $B$ and $D$ mesons. The resolution on \de\ varies between 5.6 and 14.3 \mev\ for modes with zero or one $D^{*0}$ meson in the final state, and between 11.6 and 19.5 \mev\ for modes containing two neutral $D^{*}$ mesons.

The efficiency for signal events varies from 0.5\% to 22.2\% depending on the final state (being typically in the $5\%-10\%$ range). The modes with the lowest efficiency are the ones containing one or two charged $D^{*}$ mesons.

\begin{figure*}[htb]
\begin{center}
\epsfig{file=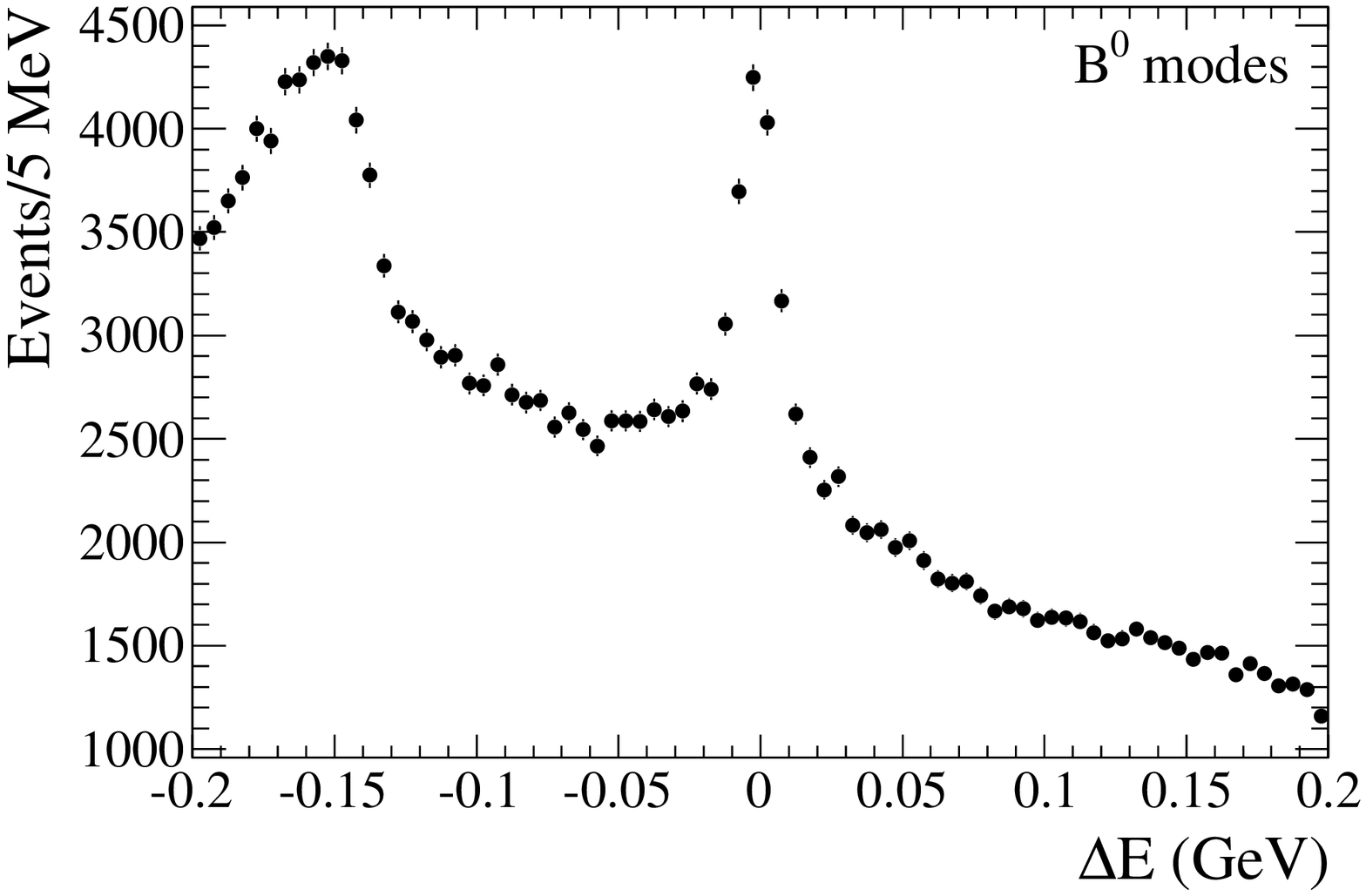,width=6.5cm} \epsfig{file=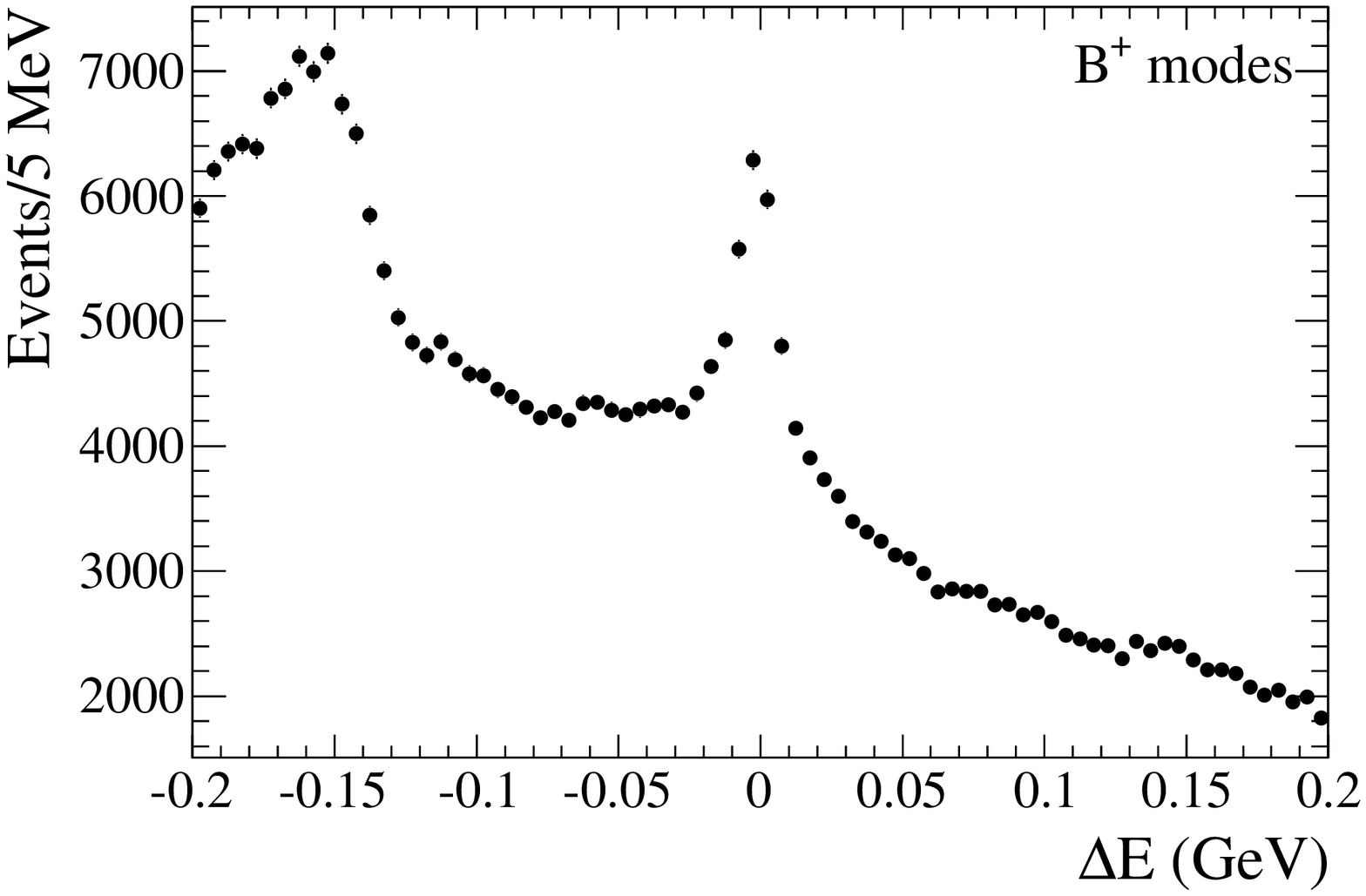,width=6.5cm}
\epsfig{file=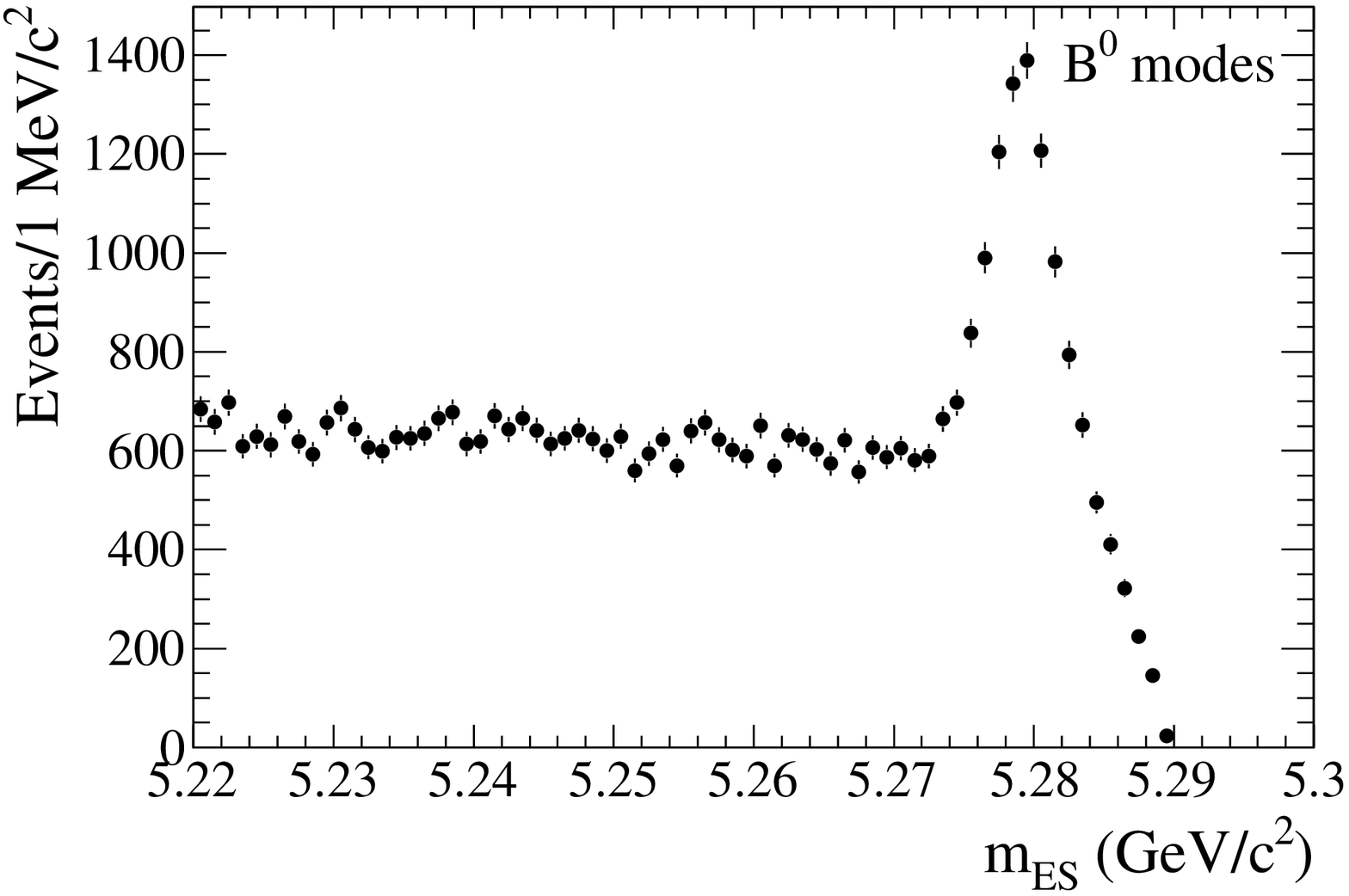,width=6.5cm} \epsfig{file=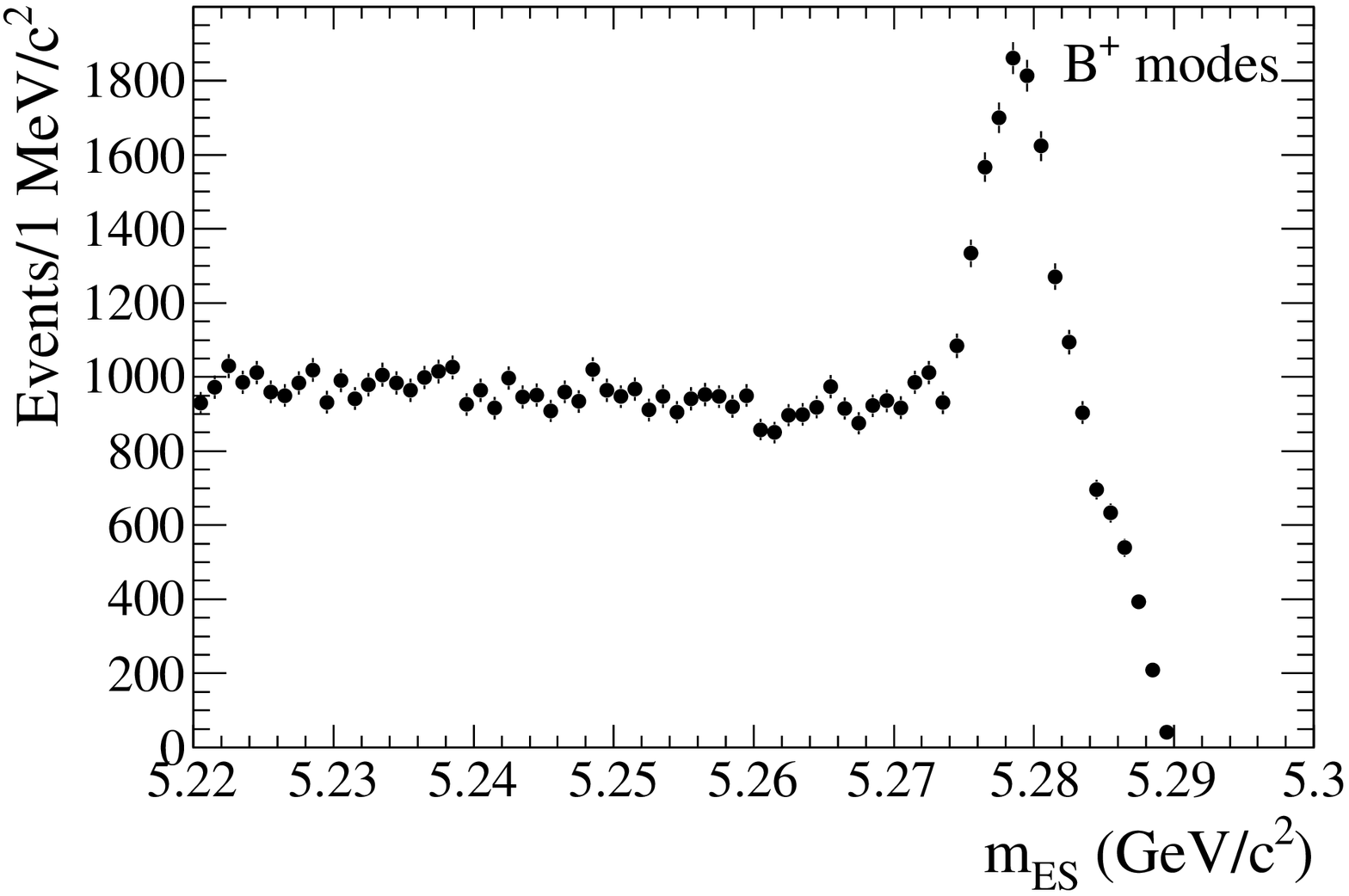,width=6.5cm}
\caption{Distributions of the \de\ variable (top plots) and of the \mes\ variable (bottom plots) for the sum of all the $\Bz\to\Dbar^{(*)}D^{(*)}K$ modes (left-hand plots) and for the sum of all the $\Bu\to\Dbar^{(*)}D^{(*)}K$  modes (right-hand plots). The \de\ distributions are shown after the complete selection but before the choice of the best candidate and for  $5.27 < \mes < 5.29 \gevcc$, and the \mes\ distributions are shown after the complete selection, including the selection on the \de\ variable.}
 \label{fig:DEmes}
\end{center}
\end{figure*}

Figure \ref{fig:DEmes} presents the \de\ and \mes\ distributions after the complete selection is applied. The \de\ distributions are presented for events in the
signal region defined by $\mes > 5.27 \gevcc$ and are shown without
applying the best candidate selection. Signal events appear in the peak near $\de \sim 0 \mev$ when reconstructed
correctly, while the peak around $-160$ \mev\ is due to $\Dbar^* D K$ and $\Dbar D^* K$
decays reconstructed as $\Dbar D K$, and to $\Dbar^* \Dstar K$ decays reconstructed as
$\Dbar^* D K$ or $\Dbar D^* K$. Both \de\ and \mes\ distributions show a clear excess of events in the signal region.

\section{Fits of the data distributions}
\label{sec:fit}

\noindent
We present the fits used to extract the branching fractions. For each mode, we fit the
\mes\ distribution between $5.22$ and $5.30 \gevcc$ to get the signal yield. The data samples corresponding to each $B$ decay mode are disjoint and the fits are performed independently for each mode.
According to their physical origin, four categories of events with differently shaped \mes\
distributions are separately considered:
\DDK\ signal events, cross-feed events, combinatorial background events, and peaking background events.
The total probability density function (PDF) is a sum of these contributions.
Event yields are obtained from extended maximum likelihood unbinned fits.

\subsection{Signal contribution}

\noindent
The shape of the signal is determined from fits to the \mes\ distributions of signal MC samples. A Crystal Ball function~\cite{ref:cb} (Gaussian modified to include a
power-law tail on the low side of the peak), $\mathcal{P}_{\mathrm{S}}(\mes; m_{\mathrm{S}}, \sigma_{\mathrm{S}},
\alpha_{\mathrm{S}}, n_{\mathrm{S}})$, is used to describe the signal (see Eq.~(\ref{eq:signal}) in the Appendix~\ref{ref:app}).
The parameters of this PDF are $m_{\mathrm{S}}$ and $\sigma_{\mathrm{S}}$, the
mean and the width of the Gaussian part, and $\alpha_{\mathrm{S}}$ and $n_{\mathrm{S}}$, the parameters of the
tail part. The signal yield, $N_{\mathrm{S}}$, is determined from the fit to the data.

\subsection{Cross-feed contribution}

\noindent
We call ``cross feed" the events from all of the
\DDK\ modes, except the one we reconstruct, that pass the complete selection and that are reconstructed in
the given mode. The cross-feed events are a non-negligible part of the \mes\ peak in some of the modes, and the signal event yield must be corrected for these cross-feed events.

We observe from the analysis of simulated samples that most of the cross feed originates from the combination of an unrelated soft $\pi^0$ or $\gamma$ with the $D^0$ decayed from the \Dstarp to form a wrong \Dstarz candidate. The cross-feed
proportion is often in the 10\% range relative to the signal yield but can be comparable or larger than the signal contribution, especially for modes containing $\Dbar^{(*)0} D^{*0}$ in the final state. To account for the cross-feed events, an iterative procedure, described in Sec.~\ref{sec:iter}, is used to extract the signal yields and the branching fractions.

Cross-feed distributions for modes containing no
$D^{*0}$ meson can be described by a Gaussian function $\mathcal{P}_{\mathrm{CF}}^{\mathrm{peaking}}(\mes; m_{\mathrm{CF}}, \sigma_{\mathrm{CF}})$ for the peaking part
where $m_{\mathrm{CF}}$ and $\sigma_{\mathrm{CF}}$ are the mean and the width of the peaking component [Eq.~(\ref{eq:CFp1})].
For modes containing at least one neutral
$D^{*}$ meson, the peaking component is described by a function $\mathcal{P}_{\mathrm{CF}}^{\prime\ \mathrm{peaking}}(\mes; m_{\mathrm{CF}}, \sigma_{\mathrm{CF}}, t_{\mathrm{CF}})$ which is able to model the tail at low mass [Eq.~(\ref{eq:CFp2})]. The parameters
$m_{\mathrm{CF}}$ and $\sigma_{\mathrm{CF}}$ represent the position of the maximum value and the width of the peak, and $t_{\mathrm{CF}}$ represents the tail of the function.
The nonpeaking part of the cross-feed contribution is described by an Argus function~\cite{ref:argus} $\mathcal{P}_{\mathrm{CF}}^{\mathrm{nonpeaking}}(\mes;m_{0},\xi_{\mathrm{CF}})$,
where $m_{0}$ represents the kinematic upper limit for the constrained mass and $\xi_{\mathrm{CF}}$ is the Argus shape parameter [Eq.~(\ref{eq:CFnp})].

The total PDF for cross feed events is
\begin{eqnarray}
\label{eq:CF}
&\mathcal{P}_{\mathrm{CF}}(\mes; m_{\mathrm{CF}}, \sigma_{\mathrm{CF}}, t_{\mathrm{CF}}, m_{0},\xi_{\mathrm{CF}}) =& \\
\nonumber &N_{\mathrm{CF}}^{\mathrm{peaking}} \times \mathcal{P}_{\mathrm{CF}}^{(\prime)\ \mathrm{peaking}} +
N_{\mathrm{CF}}^{\mathrm{nonpeaking}} \times \mathcal{P}_{\mathrm{CF}}^{\mathrm{nonpeaking}},&
\end{eqnarray}
where $\mathcal{P}_{\mathrm{CF}}^{(\prime)\ \mathrm{peaking}}$ represents either $\mathcal{P}_{\mathrm{CF}}^{\mathrm{peaking}}$ or $\mathcal{P}_{\mathrm{CF}}^{\prime\ \mathrm{peaking}}$ depending on the number of neutral $D^{*}$ meson in the final state.
The quantities $N_{\mathrm{CF}}^{\mathrm{peaking}}$ and $N_{\mathrm{CF}}^{\mathrm{nonpeaking}}$ are the
numbers of events in the peaking PDF and in the nonpeaking PDF,
respectively. The values of the parameters of the cross-feed PDF are determined by fitting signal MC \mes\ distributions, except for the value of $m_{0}$ which is fixed to 5.2892~\gevcc. The cross-feed yield, $N_{\mathrm{CF}} = N_{\mathrm{CF}}^{\mathrm{nonpeaking}} + N_{\mathrm{CF}}^{\mathrm{peaking}}$, is also extracted from the fit.

\subsection{Combinatorial background contribution}

\noindent
The combinatorial background events are composed of generic $B$ decays and of continuum events, which account, respectively, for about 88\% and 12\% of the total number of background events. The combinatorial background events are described by an Argus function $\mathcal{P}_{\mathrm{CB}}(\mes; m_{0}, \xi_{\mathrm{CB}})$,
where $\xi_{\mathrm{CB}}$ is the shape parameter [Eq.~(\ref{eq:CB})]. The parameter $\xi_{\mathrm{CB}}$ is free to float in the fit to the data while $m_{0}$ is fixed to 5.2892~\gevcc.
The yield for the combinatorial background, $N_{\mathrm{CB}}$, is also obtained from the data fit.

\subsection{Peaking background contribution}

\noindent
We call ``peaking background" the part of the background that is peaking in the signal region and that is not due to cross feed.
To extract the peaking background, we fit the \mes\ distributions from generic MC samples $\epem\to\qqbar$ ($q=u,d,s,c,b$) satisfying the \DDK\ selection and scale the results to the data luminosity.

The simulated distribution is fitted with an Argus function describing the nonpeaking
part and a Gaussian function $\mathcal{P}_{\mathrm{PB}}(\mes; m_{\mathrm{PB}}, \sigma_{\mathrm{PB}})$ describing the peaking part,
where $m_{\mathrm{PB}}$ and $\sigma_{\mathrm{PB}}$ are the mean and width of the Gaussian [Eq.~(\ref{eq:PB})].
The parameters $m_{\mathrm{PB}}$ and $\sigma_{\mathrm{PB}}$ are free to float in the fits to the simulated events, except for modes with nonconverging fits, where $m_{\mathrm{PB}}$ is fixed to the $B$ mass. These modes are \modeiii, \modexii, \modev, \modexv, \modexviii, \modeviii, \modeix, and \modexx. The fit also returns the value of the peaking background yield, $N_{\mathrm{PB}}$, which is shown in Table~\ref{tab:result}. Only the peaking part $\mathcal{P}_{\mathrm{PB}}$ is used in the fit to the data, the nonpeaking part being included in the combinatorial background.

\subsection{Fits}
\label{sec:totalFits}

\noindent
We fit the \mes\ distribution using the PDFs for the signal, for the
cross feed, for the combinatorial background, and for the peaking background as detailed in the previous sections.
The total PDF $\mathcal{P}_\mathrm{tot}$ can be written as
\begin{eqnarray}
\mathcal{P}_\mathrm{tot} &=& N_{\mathrm{S}} \times \mathcal{P}_{\mathrm{S}}(\mes; m_{\mathrm{S}}, \sigma_{\mathrm{S}}, \alpha_{\mathrm{S}}, n_{\mathrm{S}}) \\
\nonumber &+&  N_{\mathrm{CF}} \times \mathcal{P}_{\mathrm{CF}}(\mes; m_{\mathrm{CF}}, \sigma_{\mathrm{CF}}, t_{\mathrm{CF}}, m_{0},
\xi_{\mathrm{CF}}) \\
\nonumber &+& N_{\mathrm{CB}} \times \mathcal{P}_{\mathrm{CB}}(\mes; m_{0}, \xi_{\mathrm{CB}}) \\
\nonumber &+& N_{\mathrm{PB}} \times \mathcal{P}_{\mathrm{PB}}(\mes; m_{\mathrm{PB}}, \sigma_{\mathrm{PB}}).
\end{eqnarray}
The free parameters of the fit are $N_{\mathrm{S}}$, $m_{\mathrm{S}}$, $N_{\mathrm{CB}}$, and $\xi_{\mathrm{CB}}$.
All other parameters, except $m_{0}$, are fixed to the values obtained from the simulation. For modes with low signal statistics in the data, namely \modeii, \modevi\ and \modeix, we fix $m_{\mathrm{S}}$ to the value obtained from the simulation.

The free parameters are extracted by maximizing the unbinned extended likelihood
\begin{equation}
\mathcal{L} = \frac{e^{-N}N^n}{n!} \prod^n_{i=1} \mathcal{P}_\mathrm{tot},
\end{equation}
where $n$ is the number of events in the sample and $N$ is the expectation value for the total number of events.

\subsection{Iterative procedure}
\label{sec:iter}

\noindent
 Because of the presence of cross-feed events, the fit for the branching fraction for one channel uses as inputs the branching fractions from other channels. Since these branching fractions are in principle not known, we employ an iterative procedure.
In practice, we perform the complete analysis for each $B$ mode, using as a starting point the branching fractions measured by \babar\ in Ref.~\cite{ref:patrick}. We obtain new measurements of the branching fractions that we use in the next step to fix the cross-feed proportion. We repeat this procedure until the differences between the actual branching fractions and the previous ones are smaller than 2\% of the statistical uncertainty. Using this criterion, four iterations are needed. We keep the last iteration as the final result.

\subsection{Fit results}

\begin{figure*}[htb]
\begin{center}
\epsfig{file=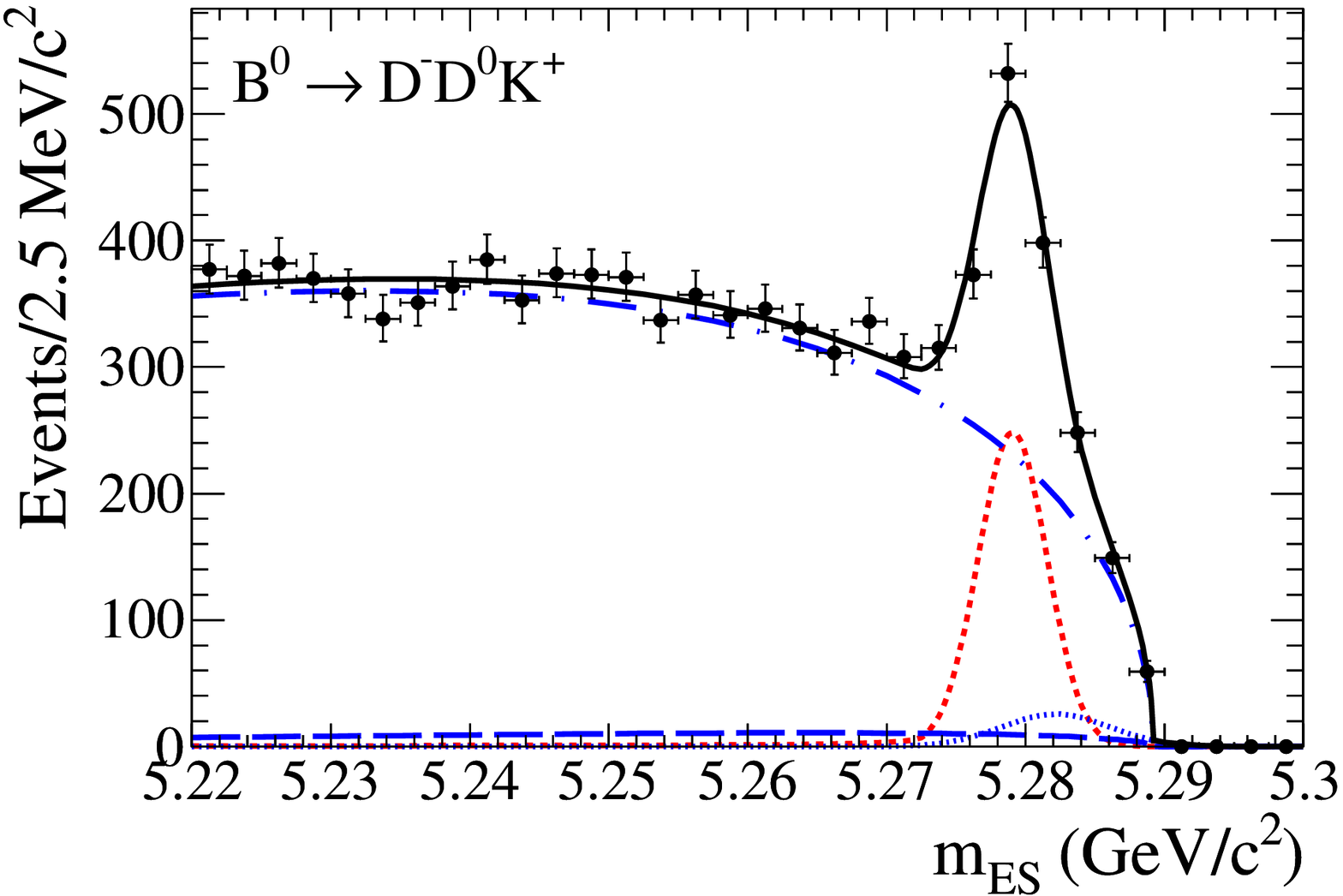,width=5.5cm}
 \epsfig{file=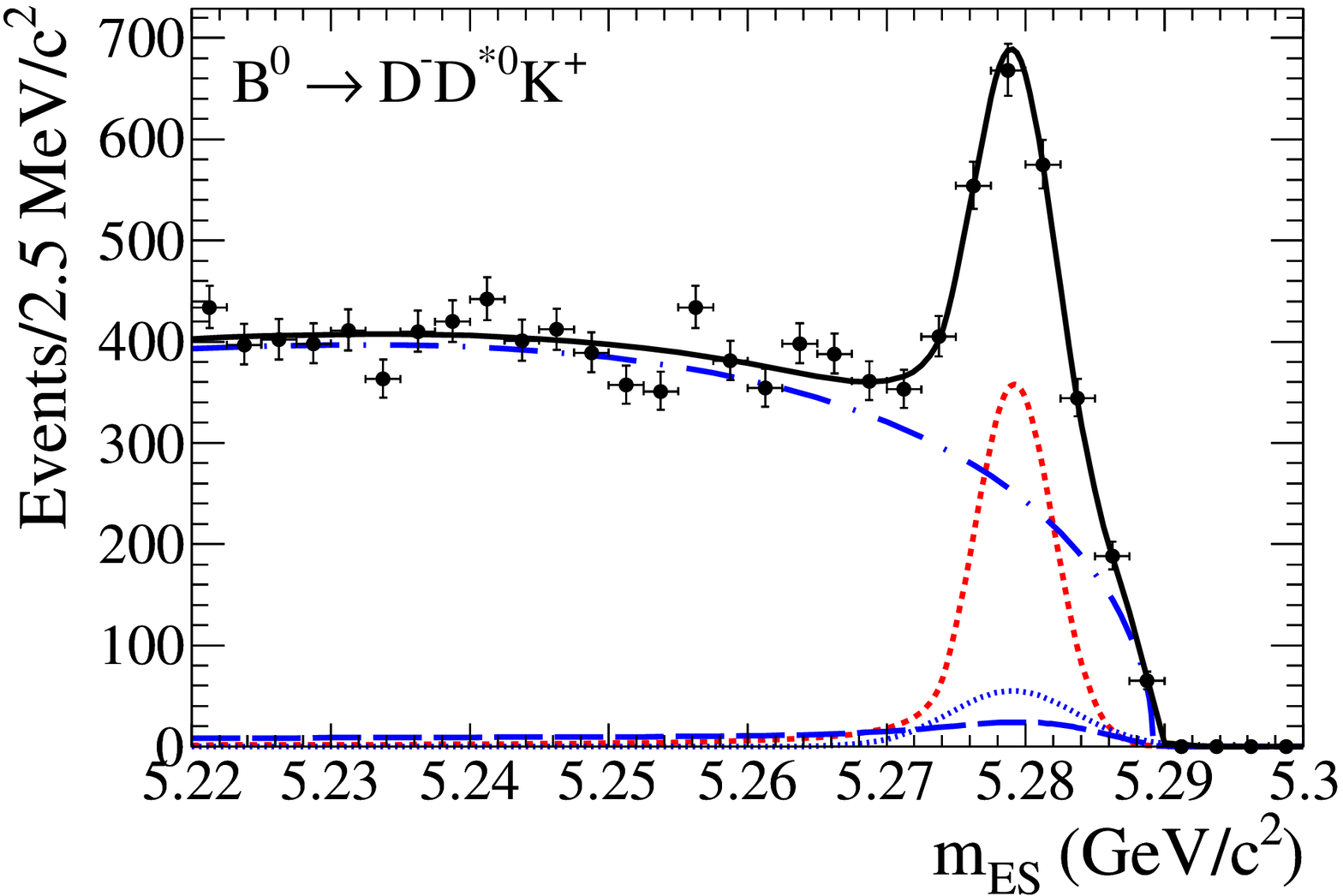,width=5.5cm}
 \epsfig{file=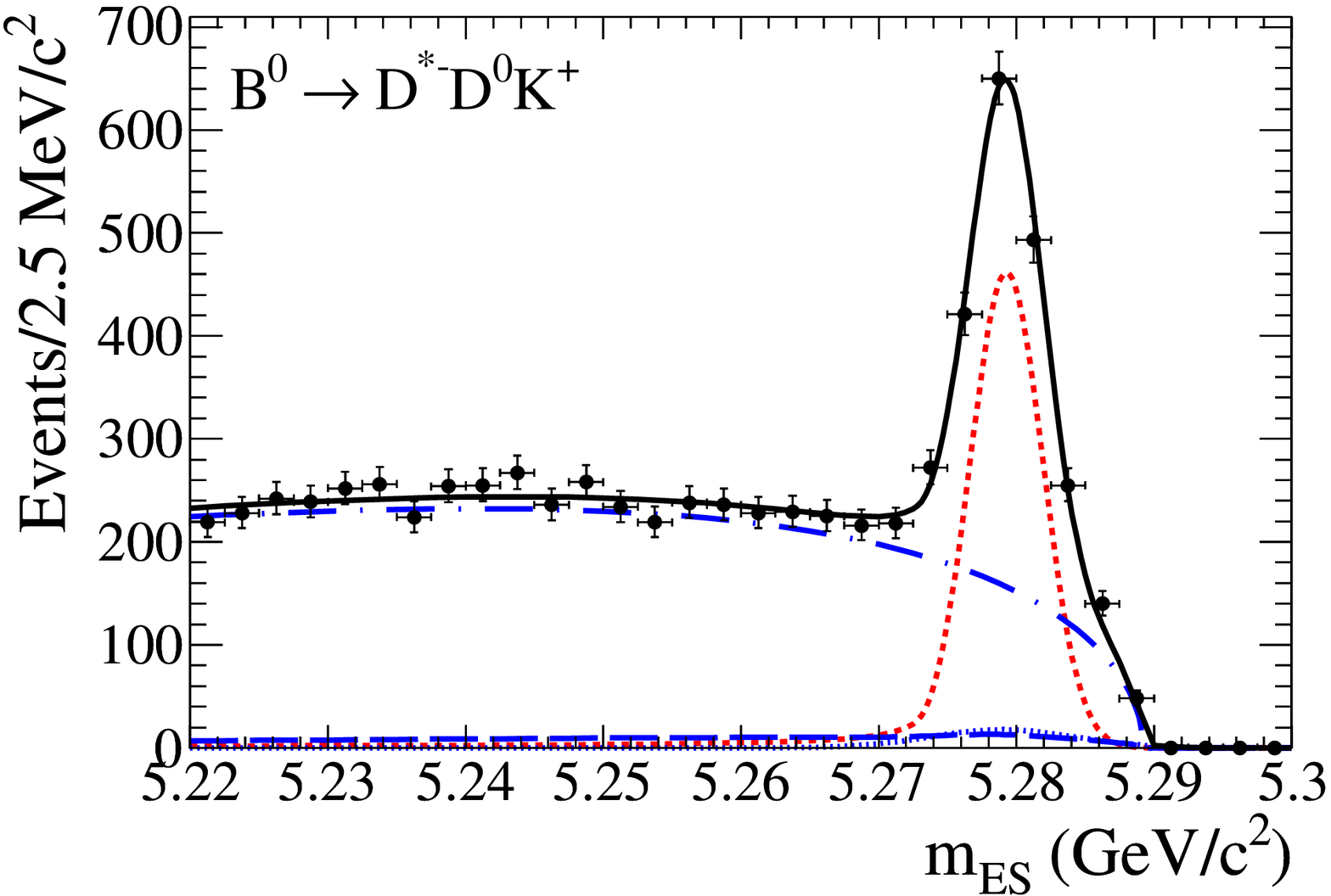,width=5.5cm}
 \epsfig{file=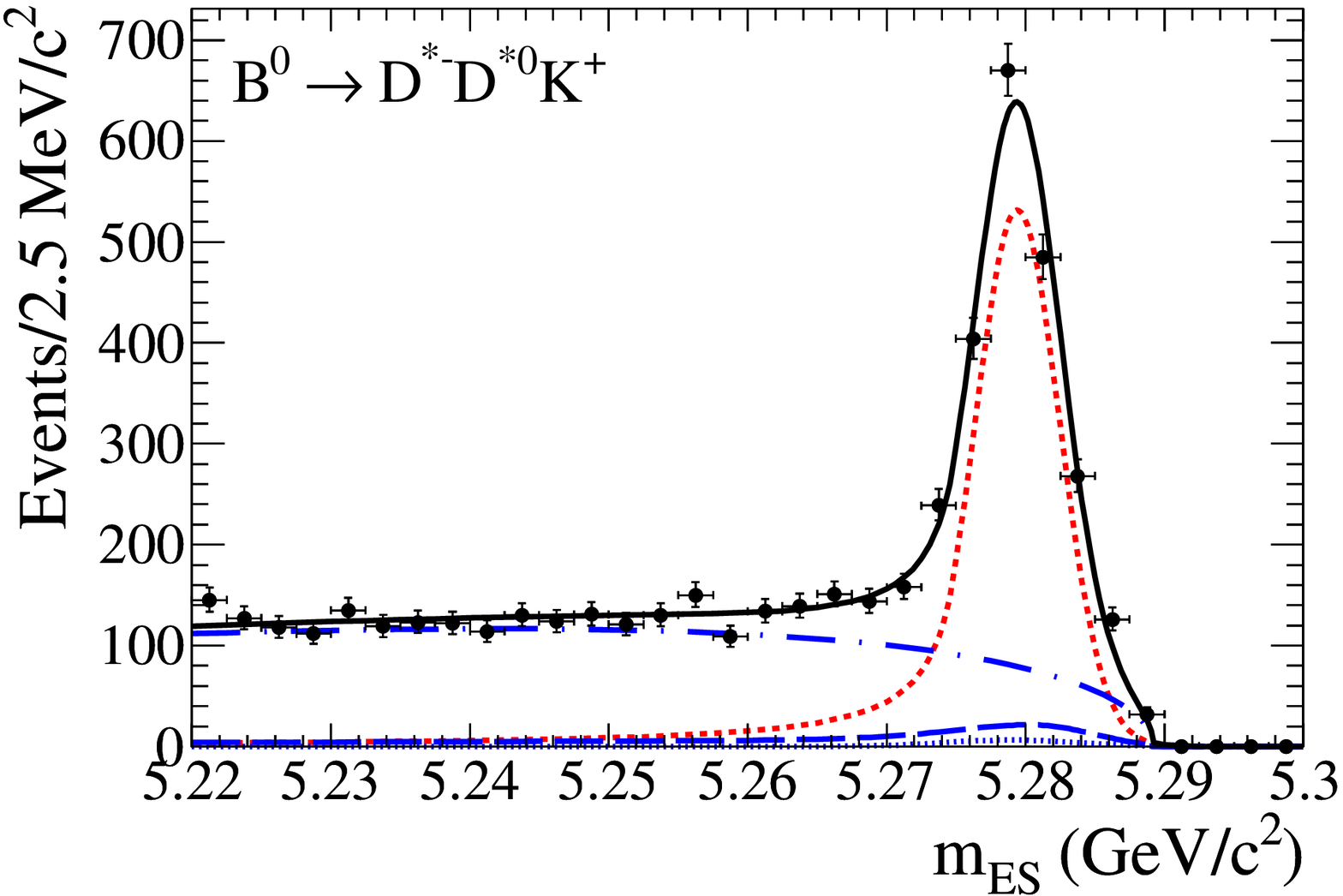,width=5.5cm}
 \epsfig{file=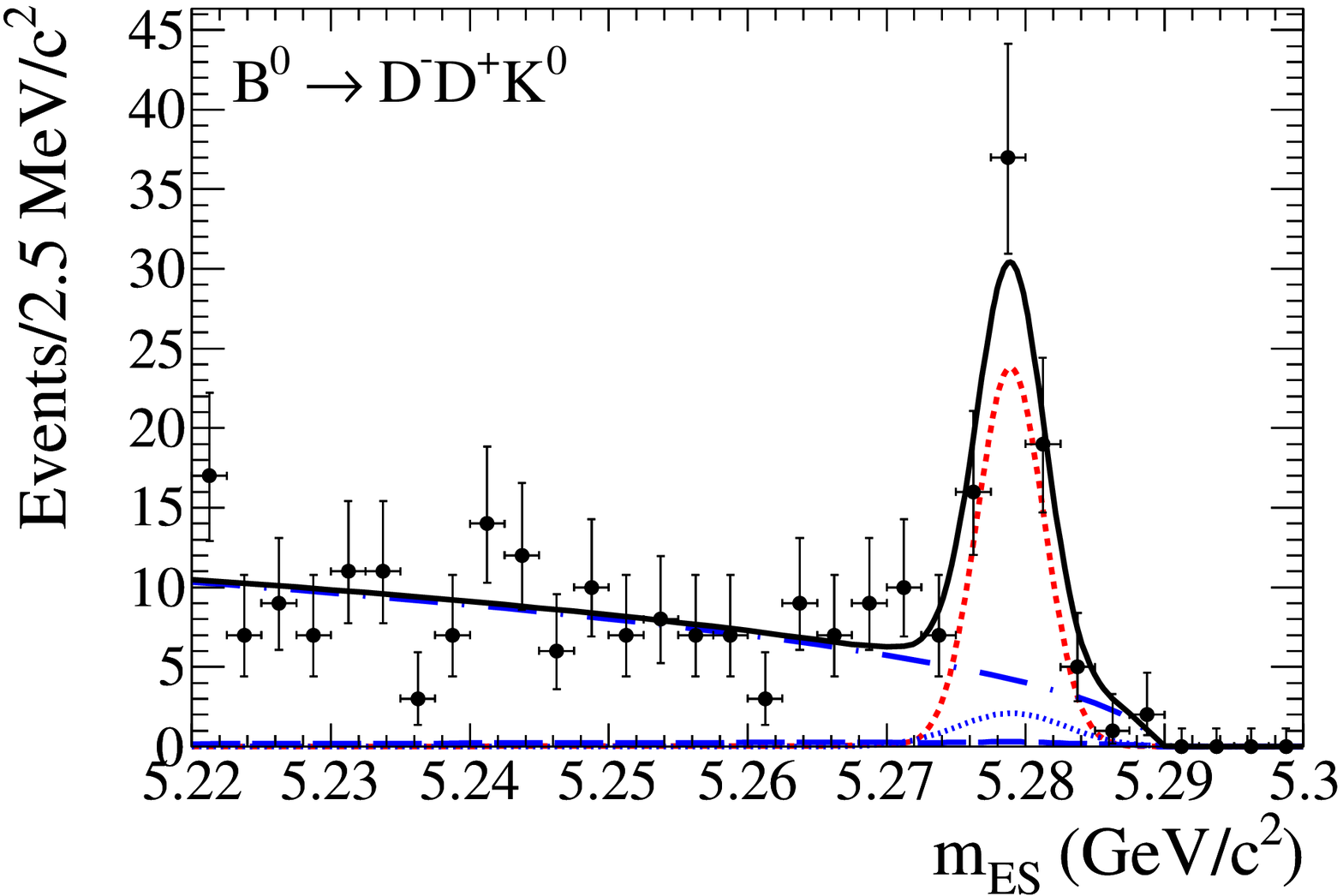,width=5.5cm}
 \epsfig{file=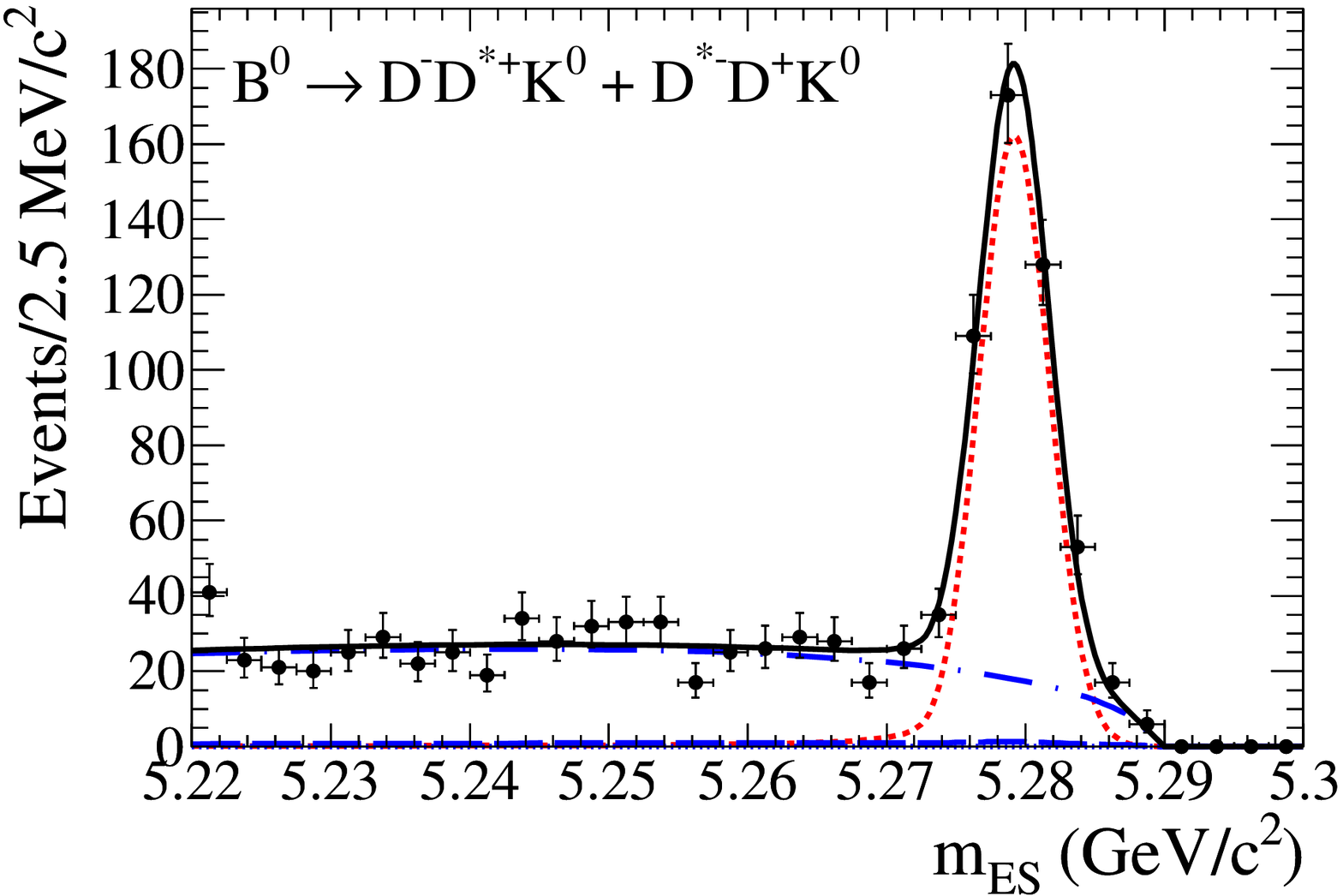,width=5.5cm}
 \epsfig{file=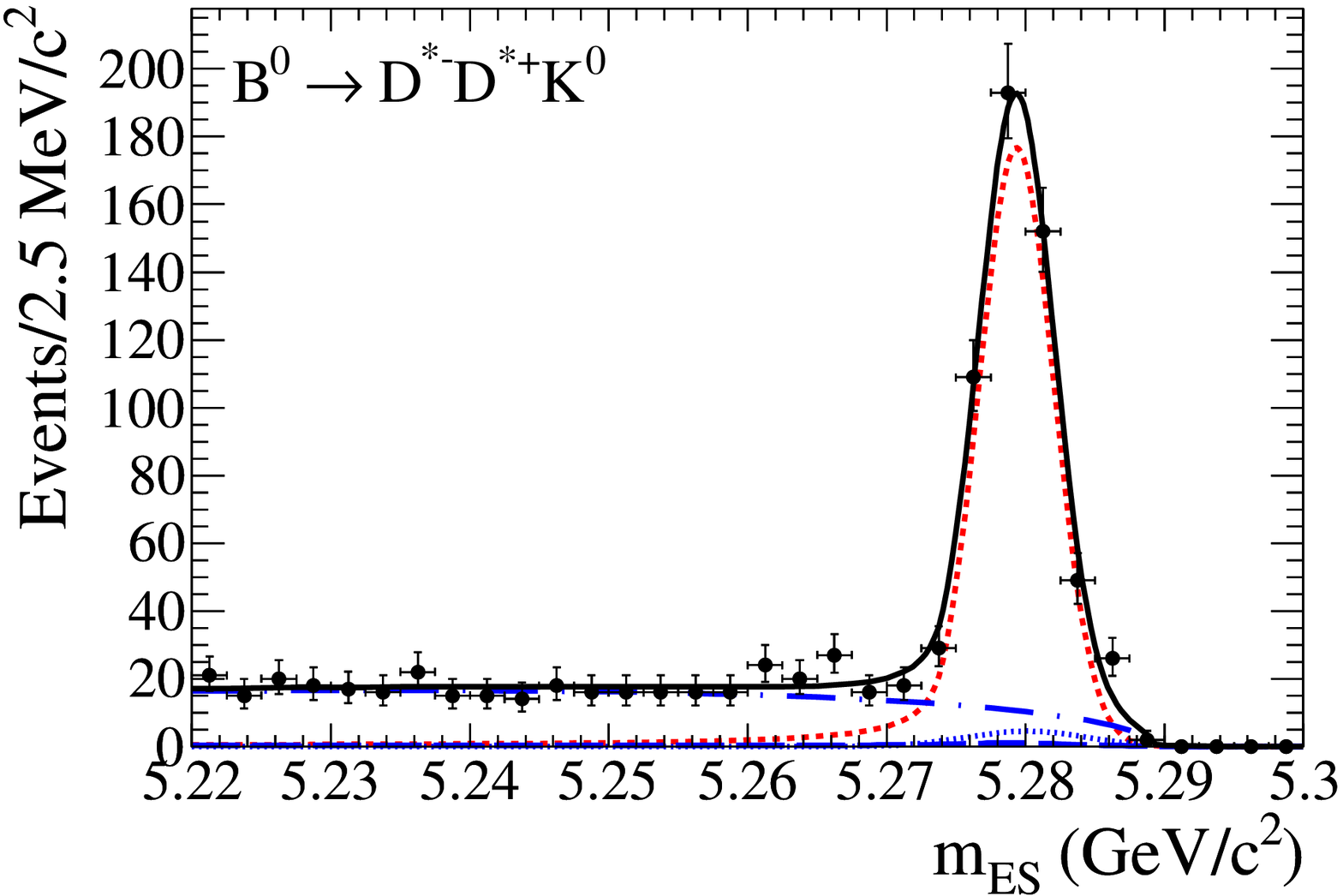,width=5.5cm}
 \epsfig{file=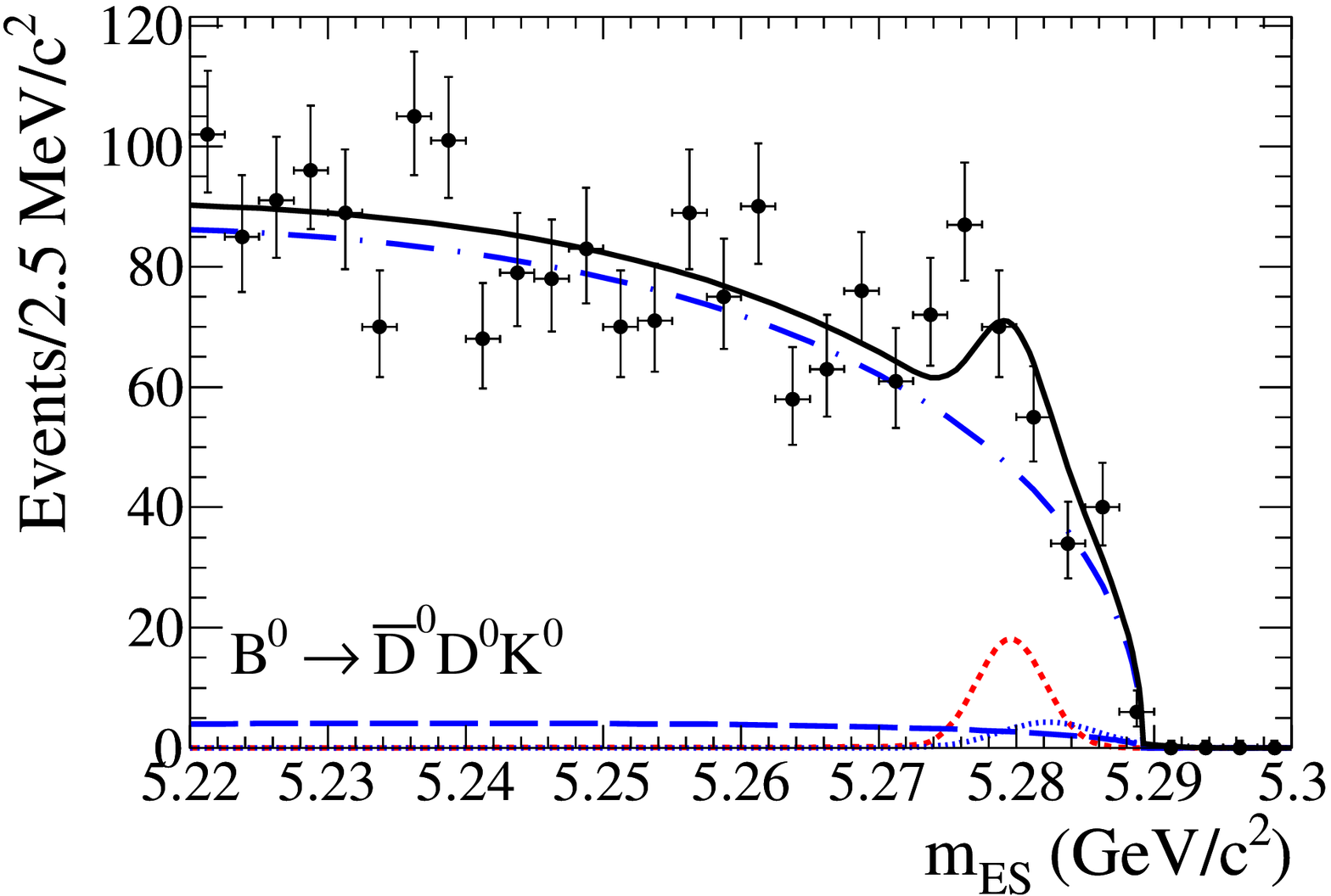,width=5.5cm}
 \epsfig{file=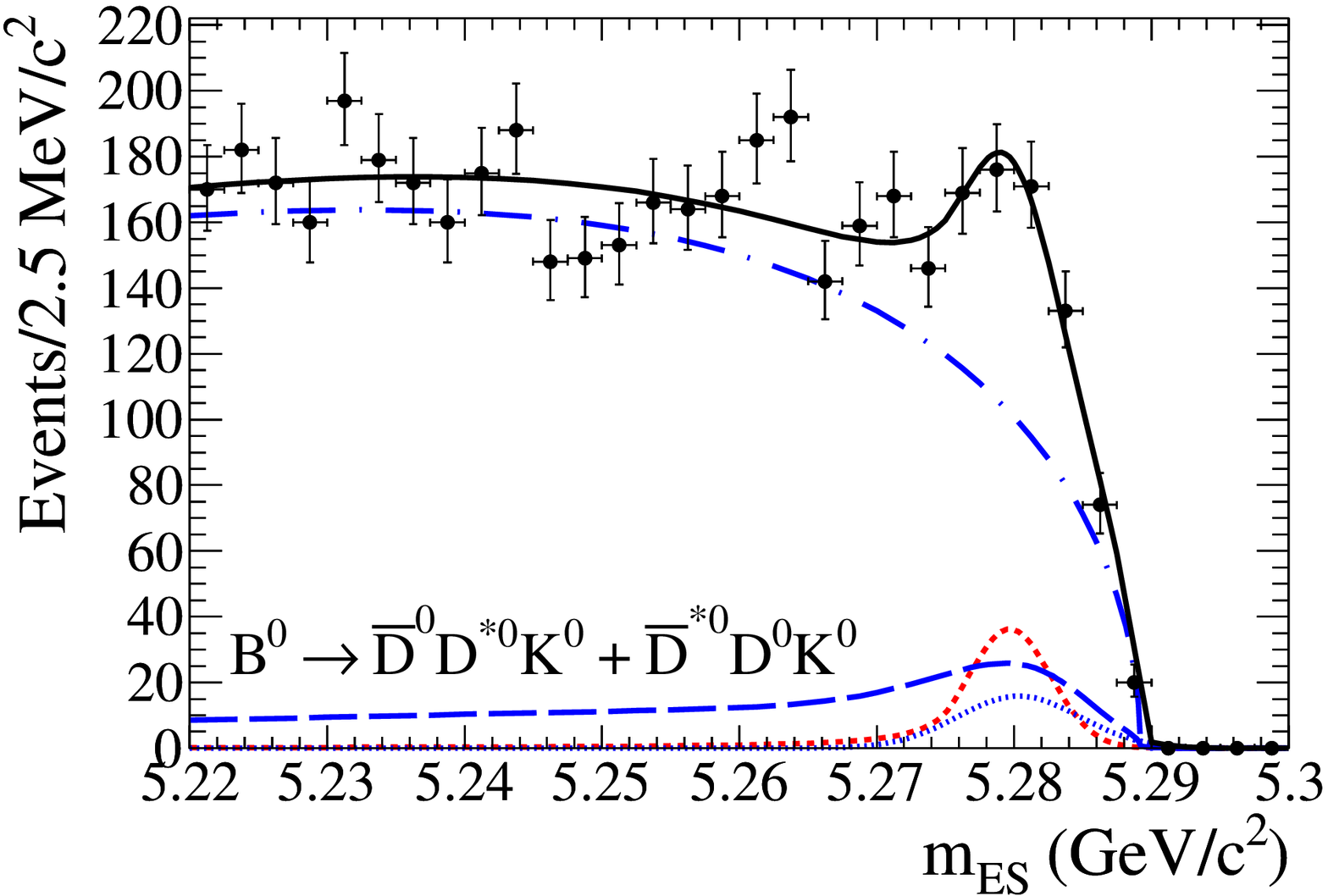,width=5.5cm}
 \epsfig{file=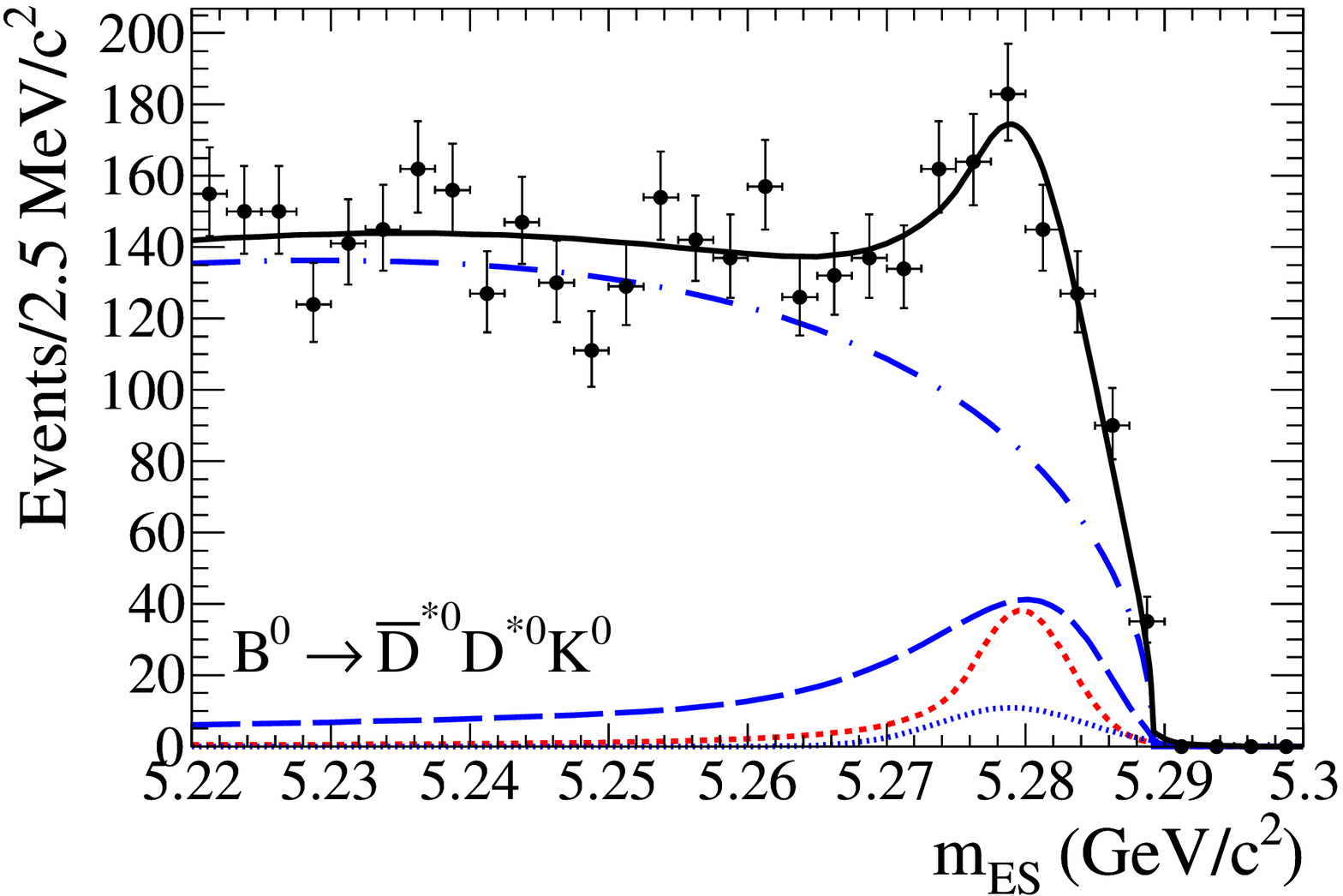,width=5.5cm}
 \caption{Fits of the \mes\ data distributions for the neutral modes, $B^0 \to \DDK$. The decay mode is indicated in the plots.
 Points with statistical errors are data events, the red dashed line represents the signal PDF, the blue long-dashed line represents the cross-feed event PDF, the blue dashed-dotted line represents the combinatorial background PDF, and the
 blue dotted line represents the peaking background PDF. The black solid line shows the total PDF. }
  \label{fig:fit1}
\end{center}
\end{figure*}

\begin{figure*}[htb]
\begin{center}
\epsfig{file=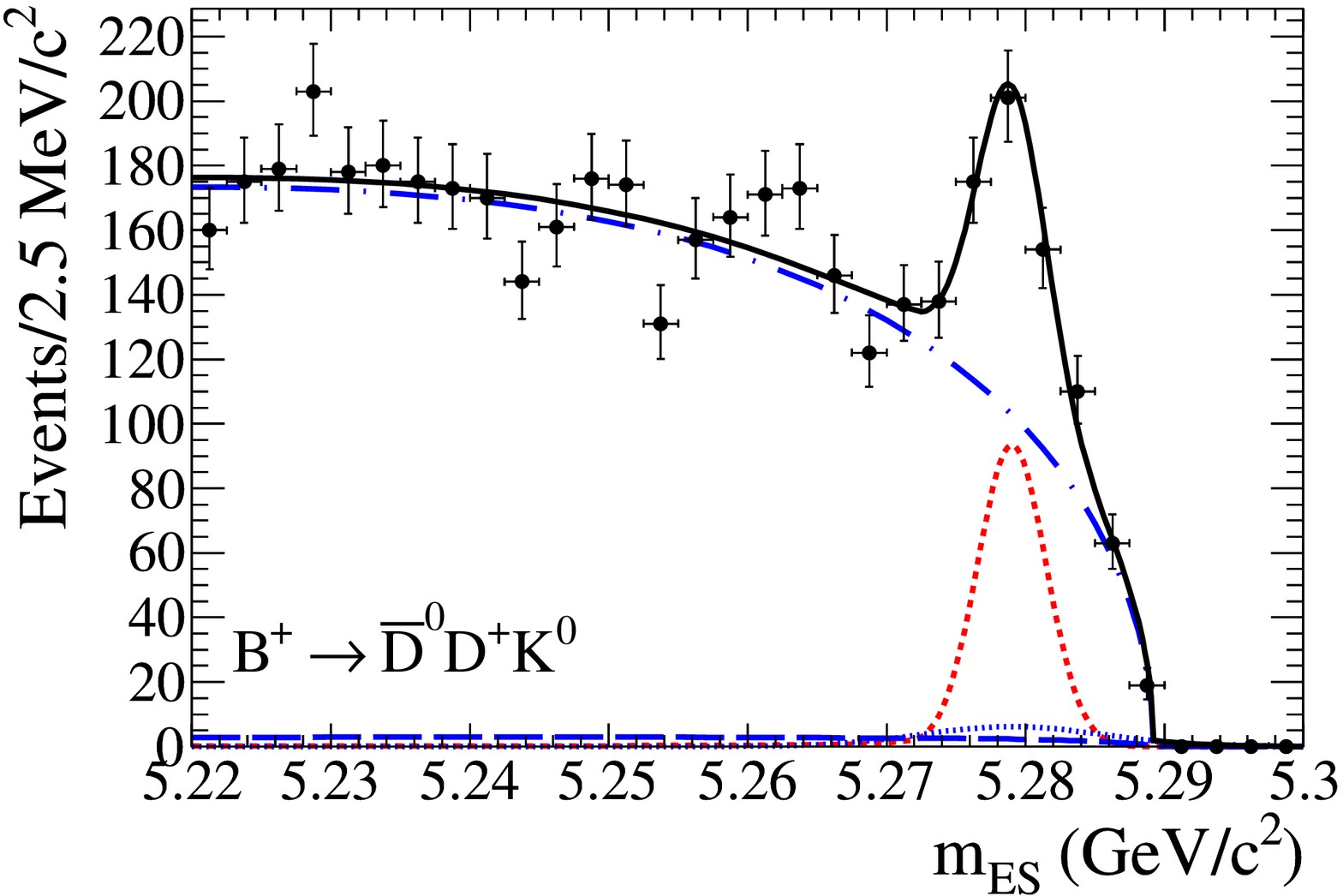,width=5.5cm}
 \epsfig{file=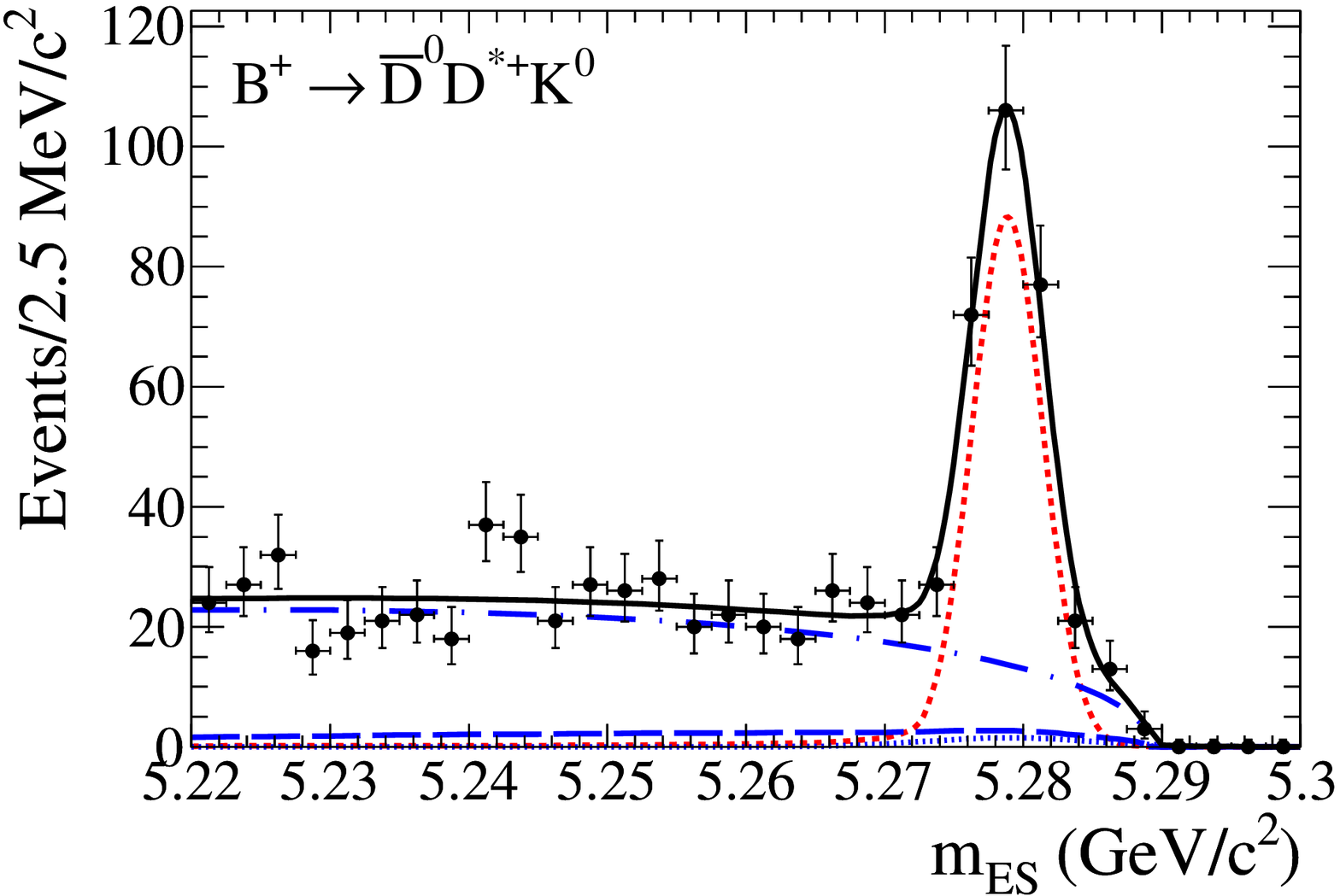,width=5.5cm}
 \epsfig{file=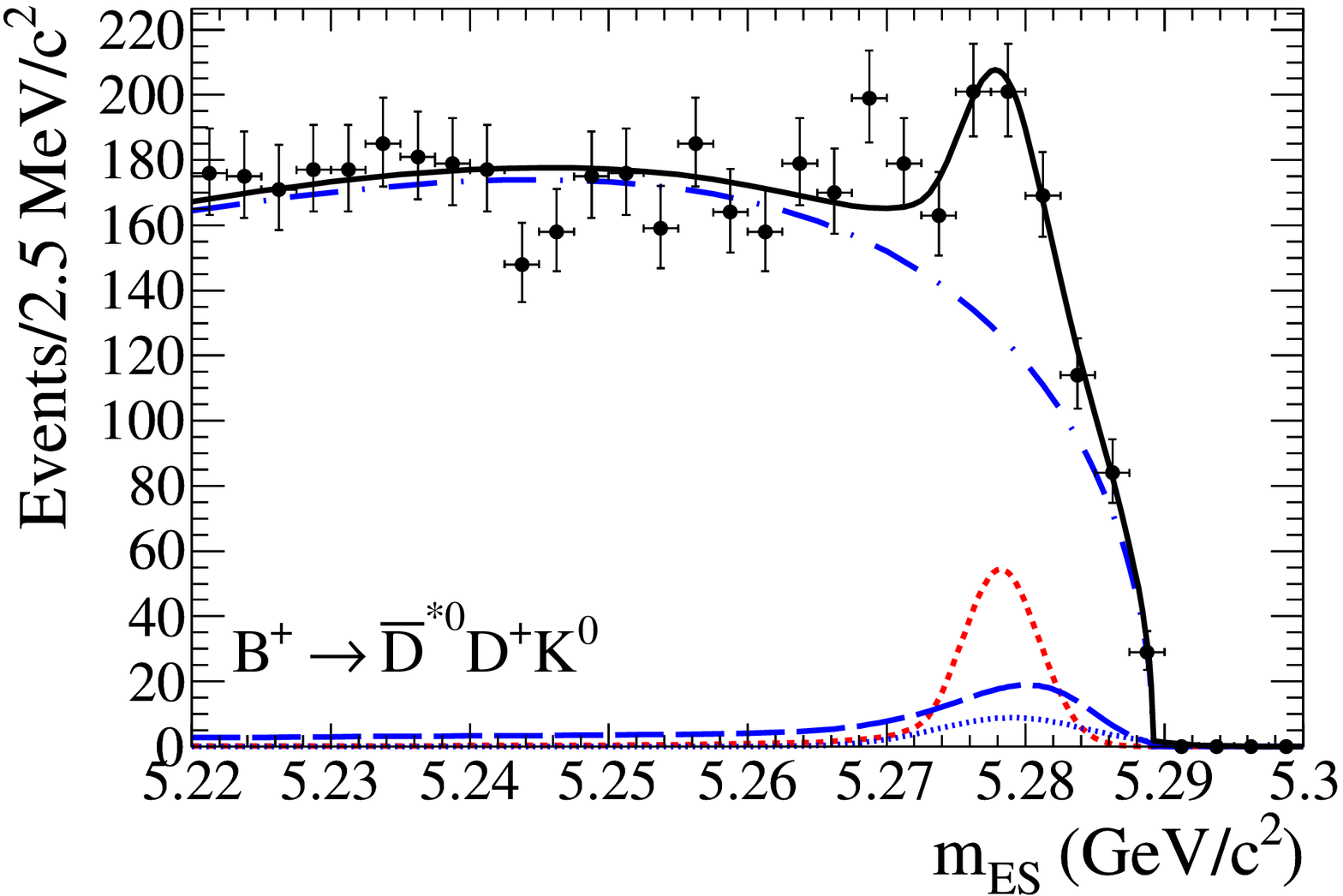,width=5.5cm}
 \epsfig{file=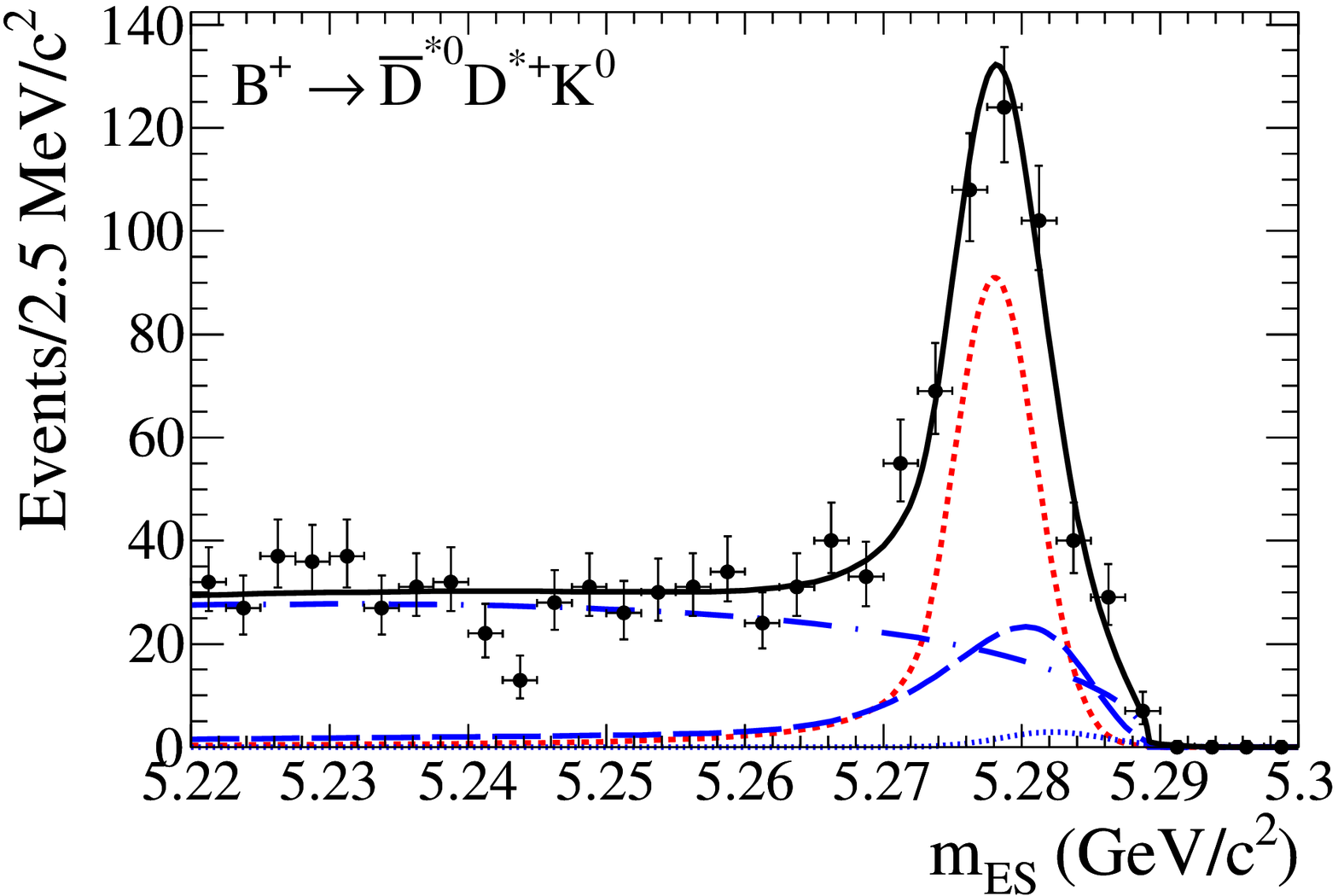,width=5.5cm}
 \epsfig{file=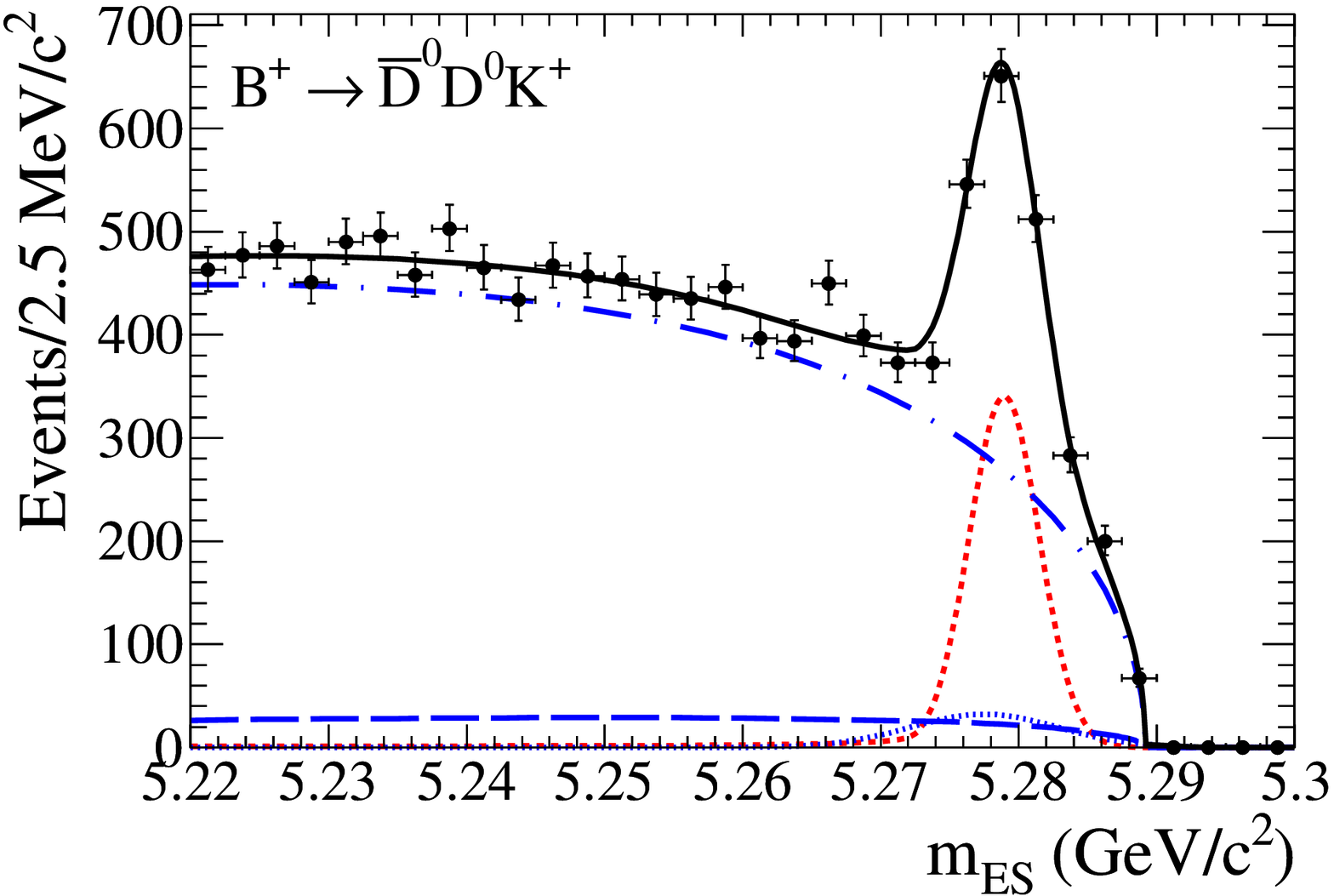,width=5.5cm}
 \epsfig{file=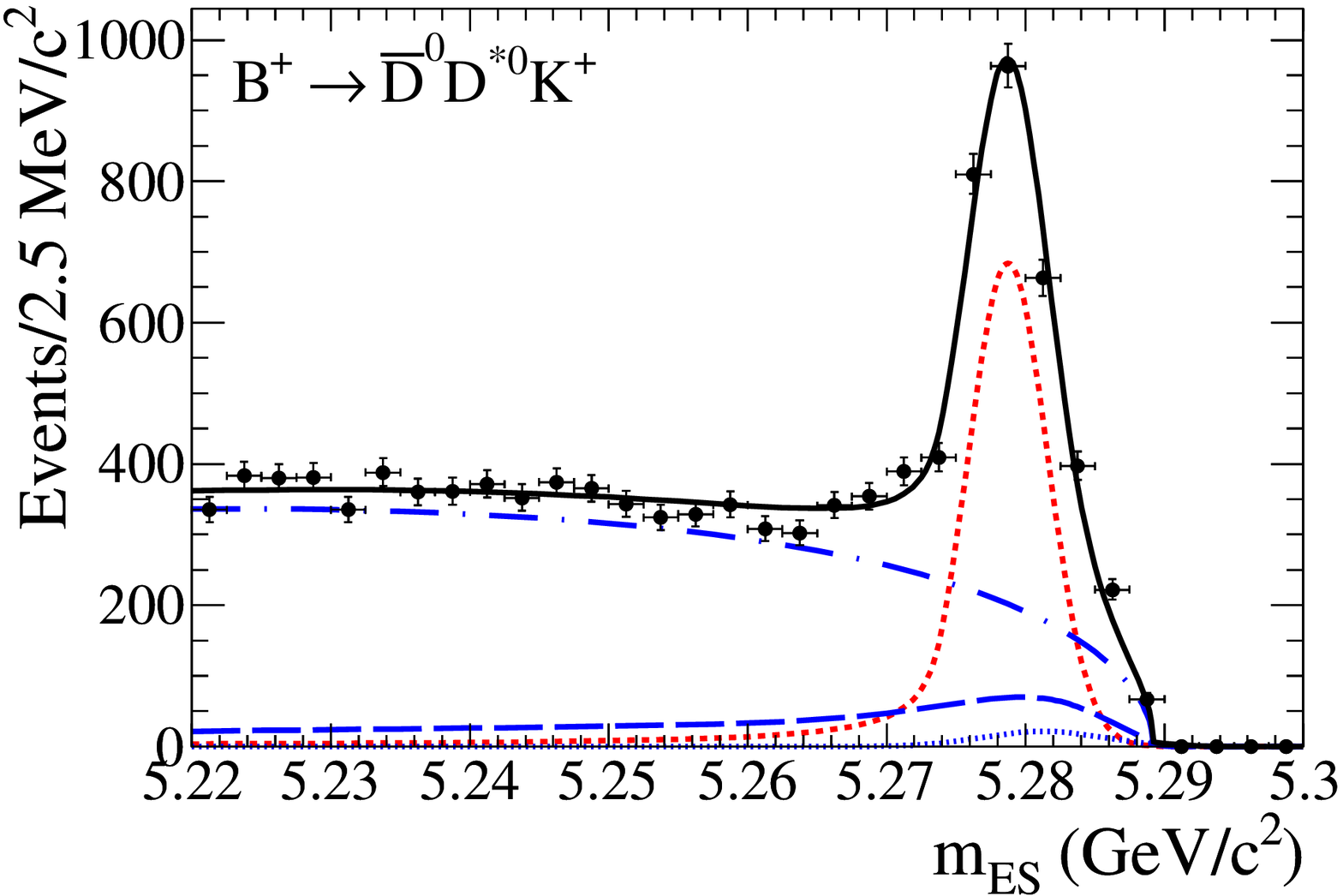,width=5.5cm}
 \epsfig{file=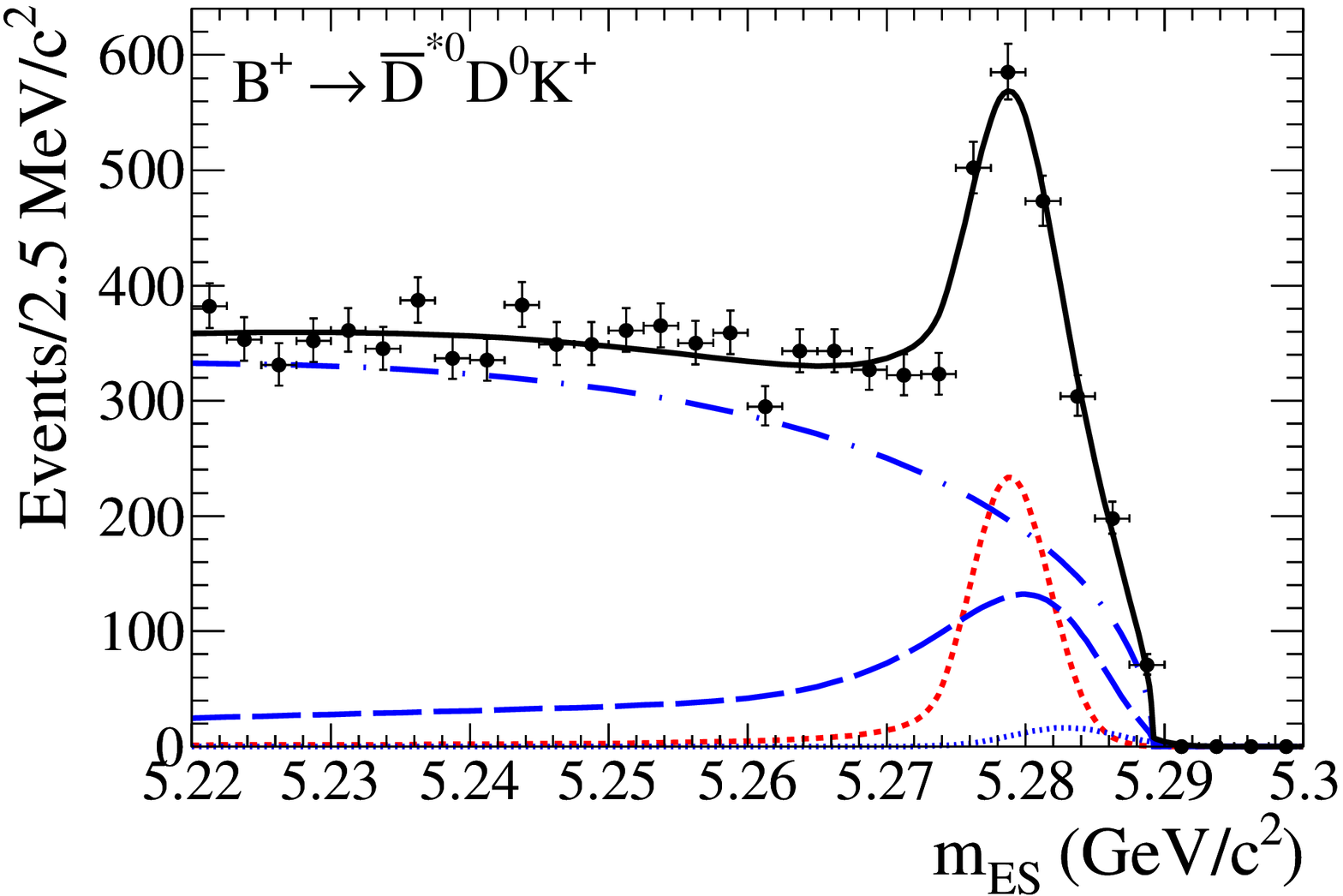,width=5.5cm}
 \epsfig{file=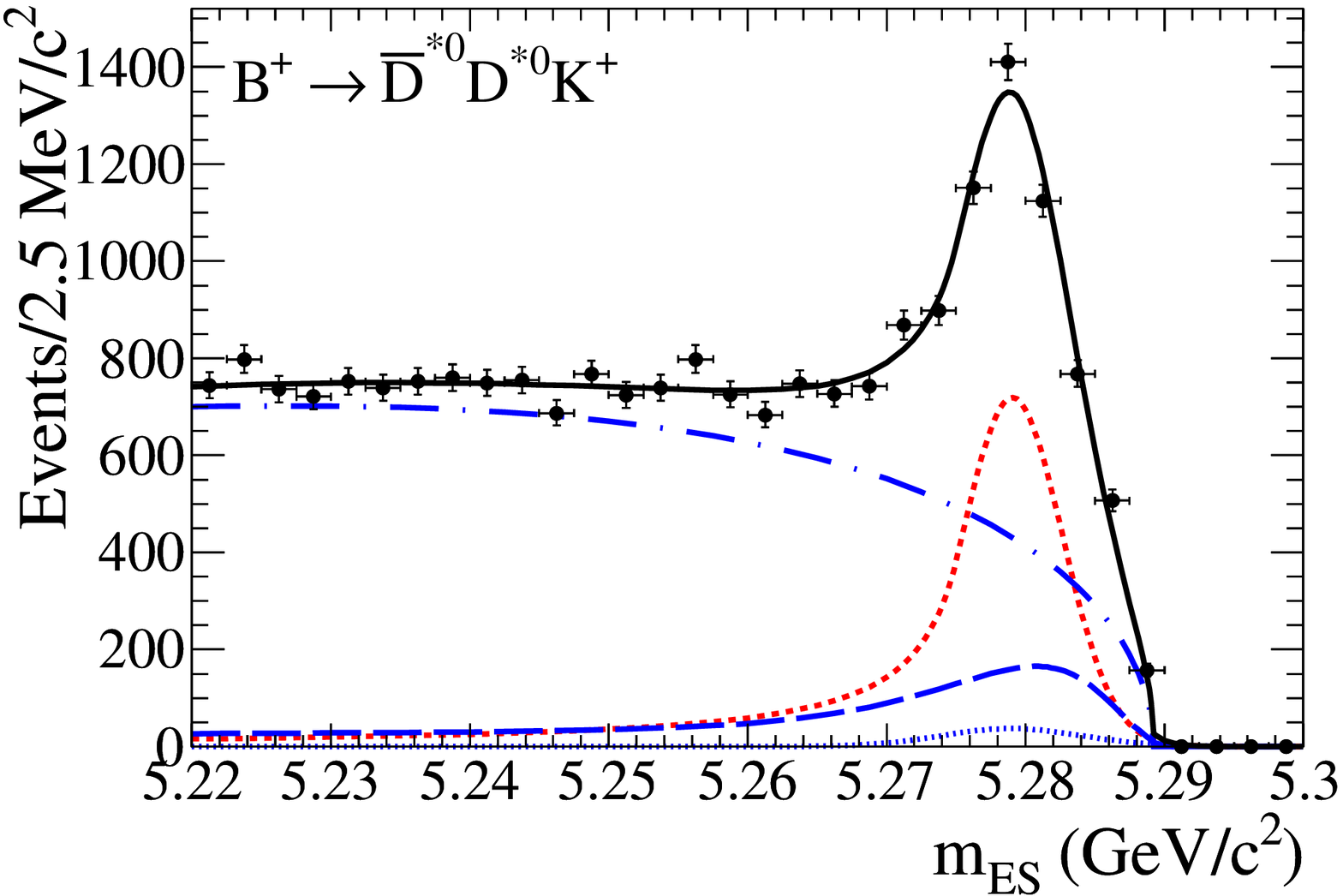,width=5.5cm}
 \epsfig{file=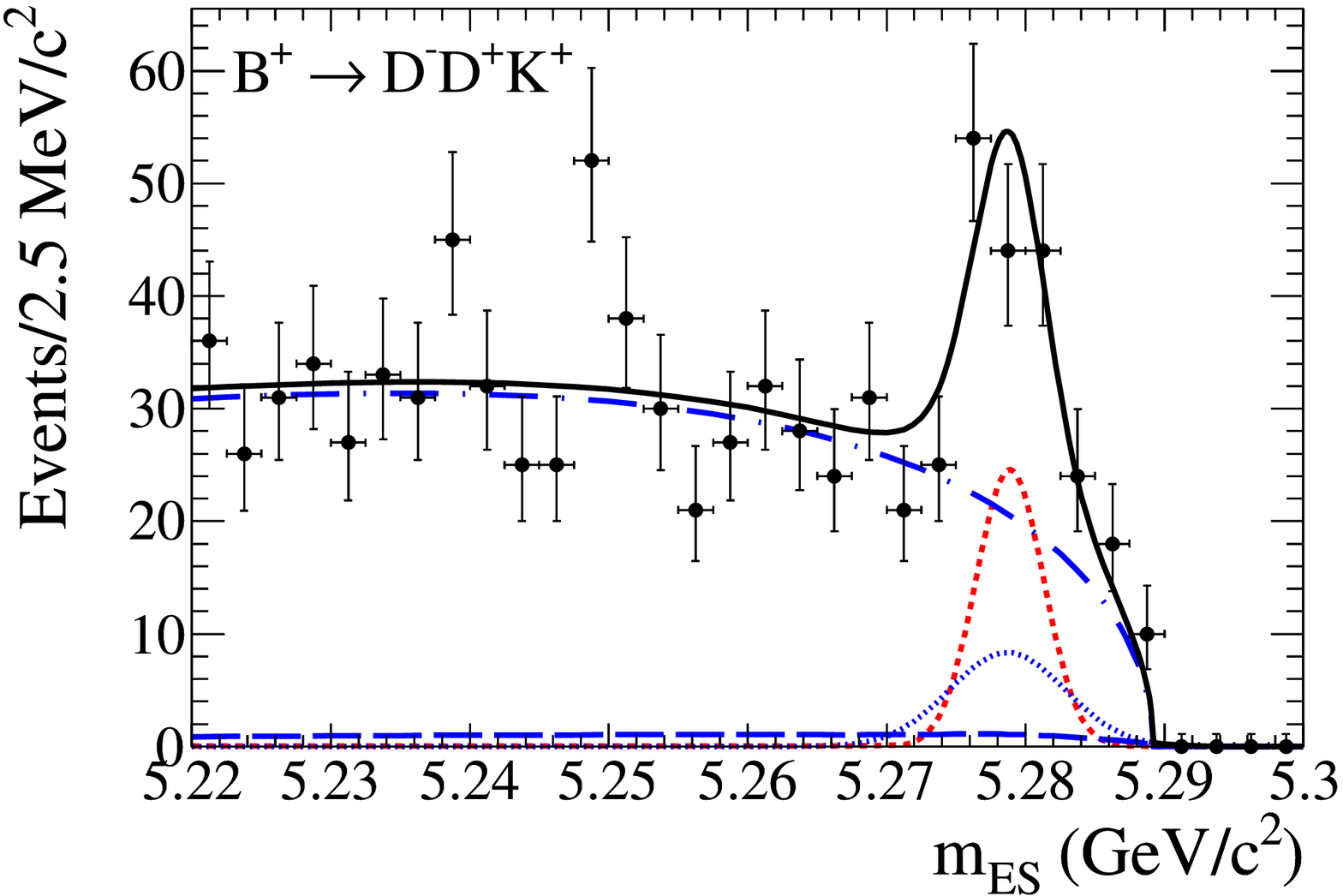,width=5.5cm}
 \epsfig{file=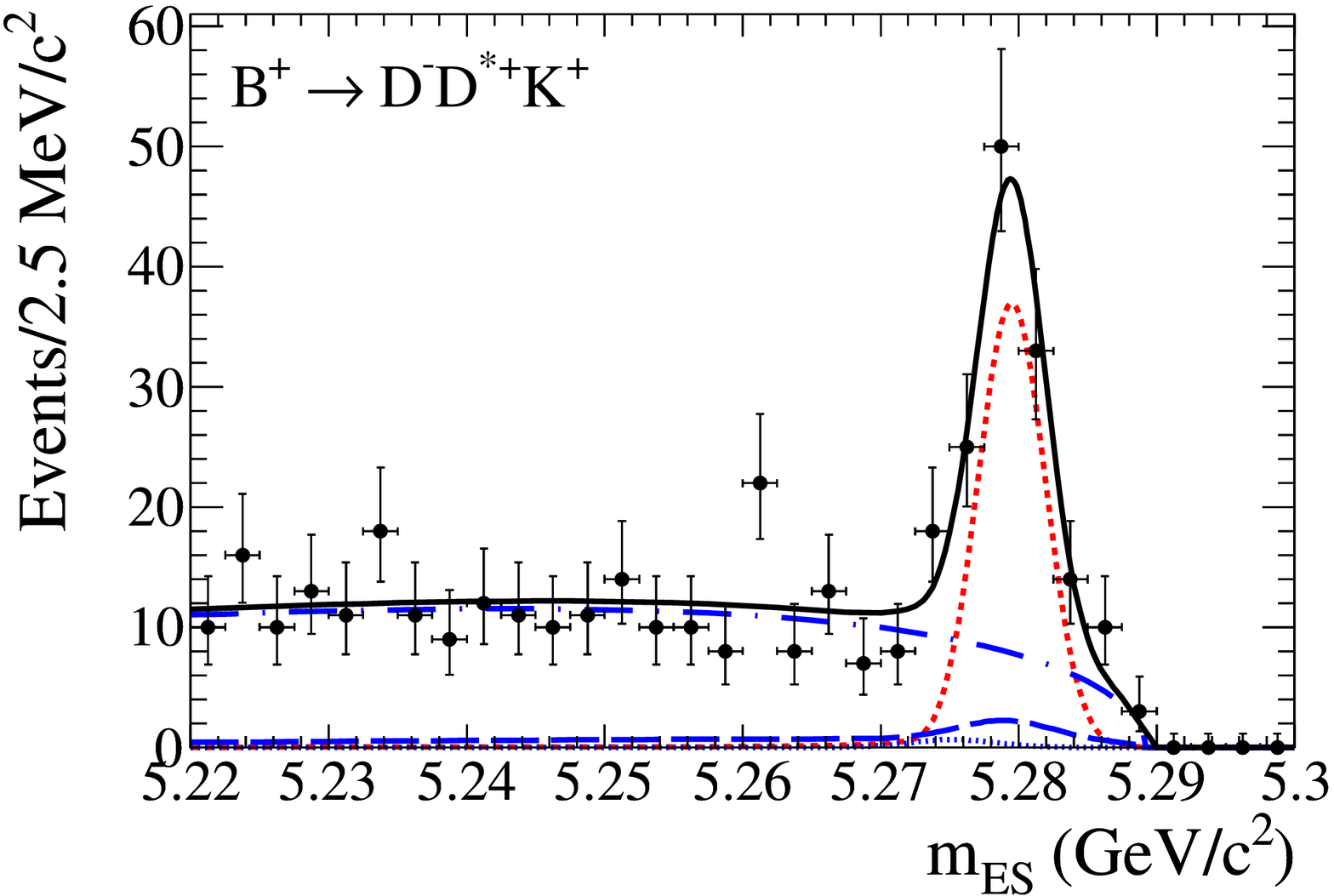,width=5.5cm}
 \epsfig{file=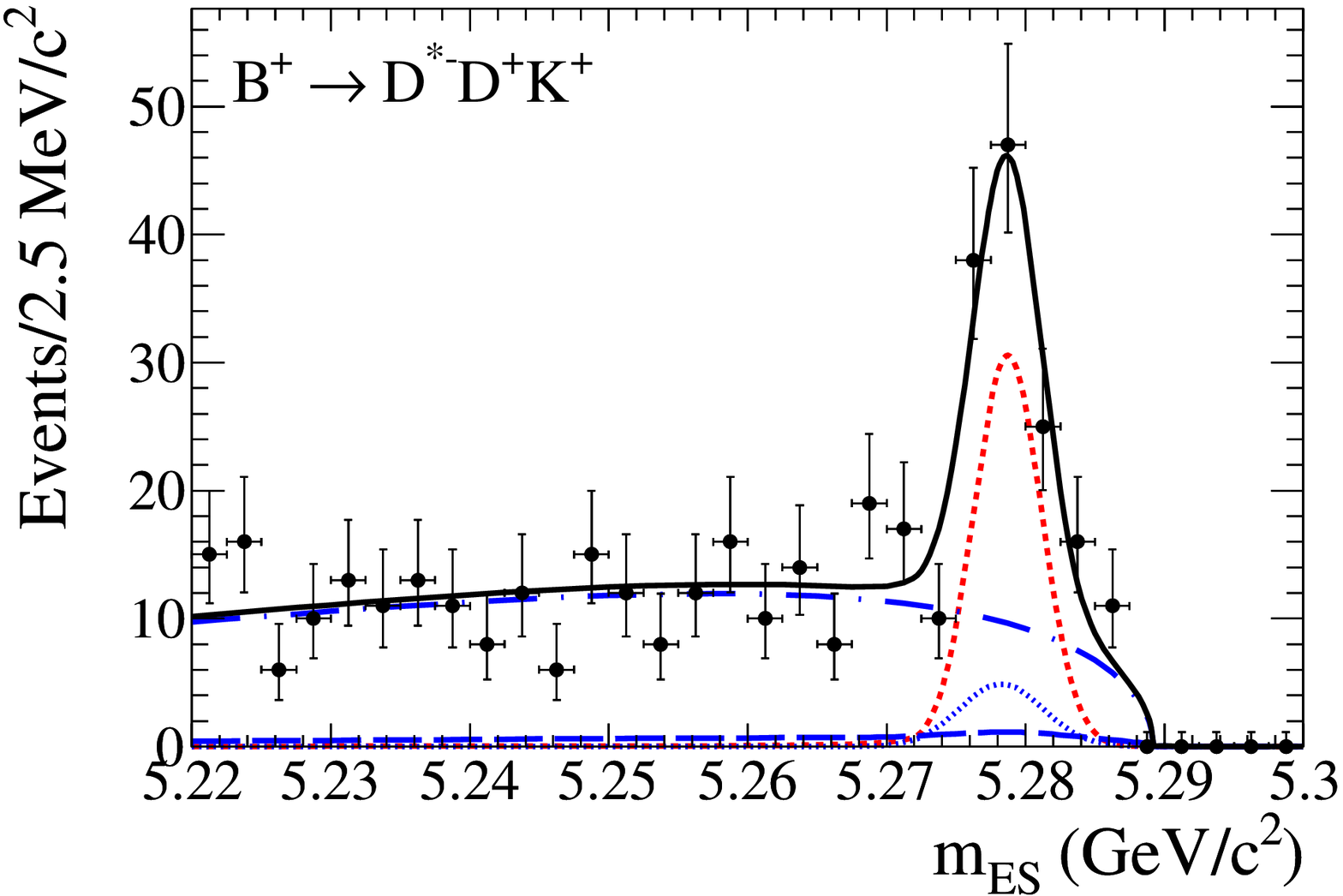,width=5.5cm}
 \epsfig{file=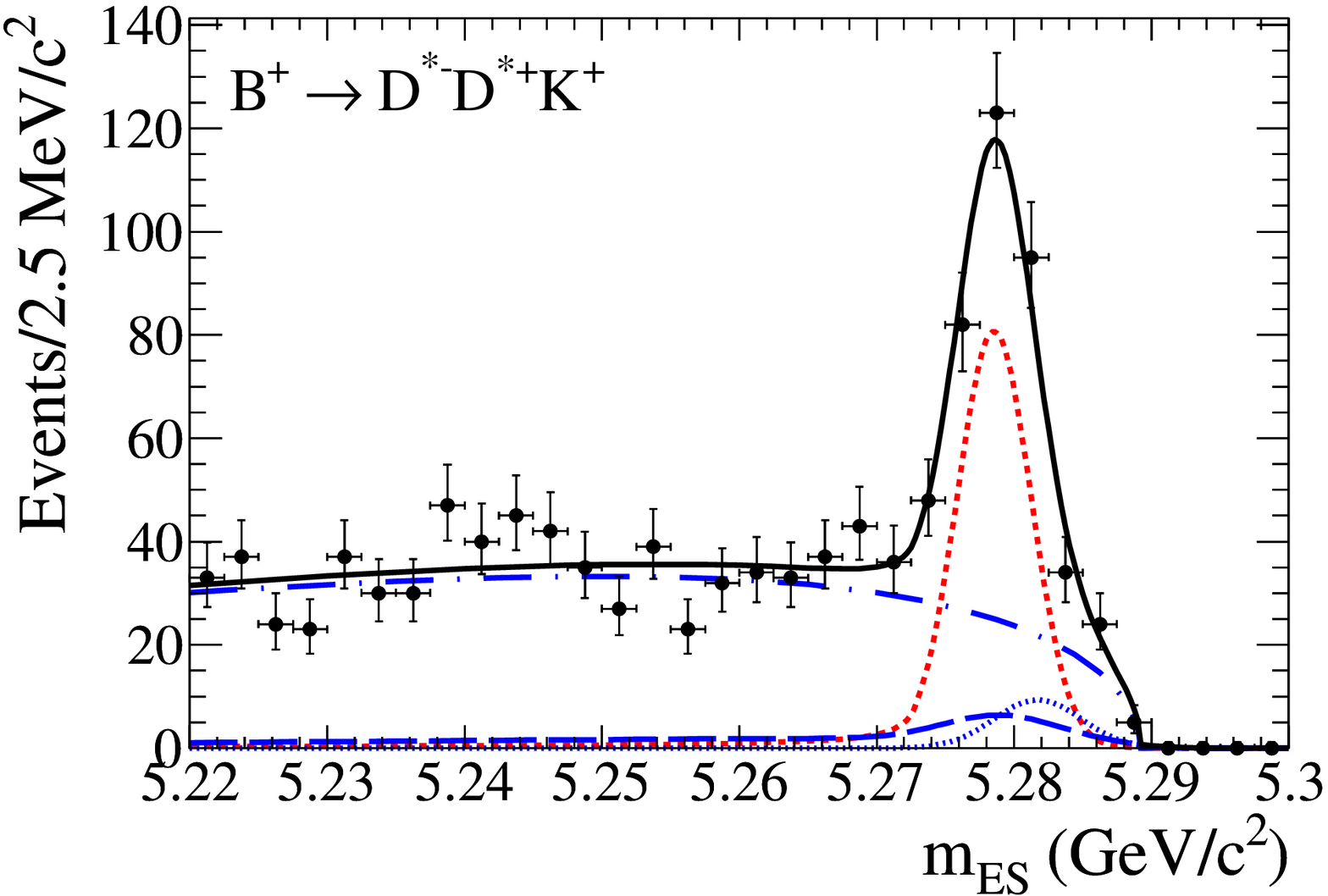,width=5.5cm}
 \caption{Fits of the \mes\ data distributions for the charged modes, $B^+ \to \DDK$. The decay mode is indicated in the plots. Points with statistical errors are data events, the red dashed line represents the signal PDF, the blue long-dashed line represents the cross-feed event PDF, the blue dashed-dotted line represents the combinatorial background PDF, and the
 blue dotted line represents the peaking background PDF. The black solid line shows the total PDF.}
  \label{fig:fit2}
\end{center}
\end{figure*}

\noindent
The results of the fits are shown in Figs.~\ref{fig:fit1} and \ref{fig:fit2}, and are displayed in Table~\ref{tab:result}.
All the fits show a good description of the data. Although we perform an unbinned fit, we can compute a $\chi^2$ value using bins of $2.5 \mevcc$ width. We observe values of $\chi^2/N_{\mathrm{dof}}$ typically close to 1, with $N_{\mathrm{dof}} = N_{\mathrm{bin}} - N_{\mathrm{float}}$, where $N_{\mathrm{bin}}$ is the number of bins and $N_{\mathrm{float}}$ is the number of floating parameters in the fit.

\section{Branching fraction measurements}

\begin{table*}[htb]
 \begin{center}
  \caption{Number of events for the signal, $N_\mathrm{S}$, for the peaking background, $N_{\mathrm{PB}}$, and for the cross feed in the signal region, $N^{\mathrm{SR}}_{\mathrm{CF}}$, and branching fractions in units of $10^{-4}$. The yields $N_\mathrm{S}$ and $N_{\mathrm{PB}}$ are defined on the whole $\mes$ range, whereas $N^{\mathrm{SR}}_{\mathrm{CF}}$ is defined for $\mes > 5.27 \gevcc$. The first uncertainties are statistical, and the second are systematic. The last column presents the significances including the systematic uncertainties.}
 \label{tab:result}
\vskip 0.2cm
 \begin{tabular*}{1.0\textwidth}{@{\extracolsep{\fill}} lrrrrr}
  \hline
  \hline
 \CellTop
Mode & $N_\mathrm{S}$~~~ & $N_{\mathrm{PB}}$~~~ & $N^{\mathrm{SR}}_{\mathrm{CF}}$ & $\BR$~~~~~~~~~~ & Significance \\
  \hline
\multicolumn{6}{c}{\CellTop $B^0$ decays through external $W$-emission amplitudes} \\
 \CellTop
\modei & 635 $\pm$ 47 & 99 $\pm$ 54 & 65 & 10.7 $\pm$ 0.7 $\pm$ 0.9 & 8.6$\sigma$ \\
\modeiv & 1116 $\pm$ 64 & 250 $\pm$ 69 & 137 & 34.6 $\pm$ 1.8 $\pm$ 3.7 & 7.6$\sigma$ \\
\modev & 1300 $\pm$ 54 & 93 $\pm$ 40 & 78 & 24.7 $\pm$ 1.0 $\pm$ 1.8 & 12.6$\sigma$ \\
\modeviii & 1883 $\pm$ 63 & 31 $\pm$ 28 & 112 & 106.0 $\pm$ 3.3 $\pm$ 8.6 & 11.4$\sigma$ \\
\hline
 \multicolumn{6}{c}{\CellTop $B^0$ decays through external+internal $W$-emission amplitudes} \\
 \CellTop
\modeiii & 58 $\pm$ 10 & 8 $\pm$ 11 & 2 & 7.5 $\pm$ 1.2 $\pm$ 1.2 & 5.1$\sigma$ \\
\modevii & 422 $\pm$ 25 & 0 $\pm$ 12 & 7 & 64.1 $\pm$ 3.6 $\pm$ 3.9 & 13.4$\sigma$ \\
\modex & 511 $\pm$ 27 & 20 $\pm$ 13 & 5 & 82.6 $\pm$ 4.3 $\pm$ 6.7 & 12.5$\sigma$ \\
\hline
 \multicolumn{6}{c}{\CellTop $B^0$ decays through internal $W$-emission amplitudes} \\
 \CellTop
\modeii & 46 $\pm$ 19 & 15 $\pm$ 19 & 19 & 2.7 $\pm$ 1.0 $\pm$ 0.5 & 2.3$\sigma$ \\
\modevi & 126 $\pm$ 39 & 70 $\pm$ 39 & 147 & 10.8 $\pm$ 3.2 $\pm$ 3.6 & 2.2$\sigma$ \\
\modeix & 170 $\pm$ 49 & 58 $\pm$ 31 & 231 & 24.0 $\pm$ 5.5 $\pm$ 6.7 & 2.2$\sigma$ \\
\hline
 \multicolumn{6}{c}{\CellTop $B^+$ decays through external $W$-emission amplitudes} \\
 \CellTop
\modexii & 237 $\pm$ 30 & 40 $\pm$ 23 & 16 & 15.5 $\pm$ 1.7 $\pm$ 1.3 & 6.6$\sigma$ \\
\modexviii & 233 $\pm$ 19 & 9 $\pm$ 10 & 17 & 38.1 $\pm$ 3.1 $\pm$ 2.3 & 10.7$\sigma$ \\
\modexv & 164 $\pm$ 37 & 48 $\pm$ 33 & 95 & 20.6 $\pm$ 3.8 $\pm$ 3.0 & 3.3$\sigma$ \\
\modexxi & 308 $\pm$ 28 & 11 $\pm$ 12 & 113 & 91.7 $\pm$ 8.3 $\pm$ 9.0 & 7.5$\sigma$ \\
\hline
 \multicolumn{6}{c}{\CellTop $B^+$ decays through external+internal $W$-emission amplitudes} \\
 \CellTop
\modexi & 901 $\pm$ 54 & 173 $\pm$ 77 & 153 & 13.1 $\pm$ 0.7 $\pm$ 1.2 & 8.6$\sigma$ \\
\modexvii & 2180 $\pm$ 74 & 92 $\pm$ 50 & 409 & 63.2 $\pm$ 1.9 $\pm$ 4.5 & 12.5$\sigma$ \\
\modexiv & 745 $\pm$ 60 & 61 $\pm$ 26 & 724 & 22.6 $\pm$ 1.6 $\pm$ 1.7 & 8.3$\sigma$ \\
\modexx & 3530 $\pm$ 141 & 186 $\pm$ 65 & 928 & 112.3 $\pm$ 3.6 $\pm$ 12.6 & 6.8$\sigma$ \\
\hline
 \multicolumn{6}{c}{\CellTop $B^+$ decays through internal $W$-emission amplitudes} \\
 \CellTop
\modexiii & 60 $\pm$ 15 & 35 $\pm$ 20 & 7 & 2.2 $\pm$ 0.5 $\pm$ 0.5 & 2.8$\sigma$ \\
\modexix & 91 $\pm$ 13 & 2 $\pm$ 7 & 10 & 6.3 $\pm$ 0.9 $\pm$ 0.6 & 6.7$\sigma$ \\
\modexvi & 75 $\pm$ 13 & 15 $\pm$ 9 & 6 & 6.0 $\pm$ 1.0 $\pm$ 0.8 & 5.1$\sigma$ \\
\modexxii & 232 $\pm$ 23 & 30 $\pm$ 14 & 31 & 13.2 $\pm$ 1.3 $\pm$ 1.2 & 7.4$\sigma$ \\
\hline
\hline
 \end{tabular*}
 \end{center}
 \end{table*}

\subsection{Method}

\noindent
In this paper, we measure the branching fractions of the 22 \DDK\ modes, including nonresonant and resonant modes. It has been shown that \DDK\ events contain resonant contributions. This was first reported by the \babar\ Collaboration in Ref.~\cite{ref:patrick} where it was observed that the three-body phase-space decay model does not give a satisfactory description of these decays.
In a subsequent study~\cite{ref:myPaper}, we showed the presence of \DsOneRes, \PsiRes\ and \XRes\ mesons in these final states. From Belle~\cite{ref:belleDsJ}, we know that the \DsJRes\ meson has a large contribution in the mode \modexi. This meson is expected to be present in \DDK\ final states containing $D^0K^+$ and $D^+K^0$, as well as in the final states containing $D^{*0}K^+$ and $D^{*+}K^0$, since it was recently seen decaying to $D^*K$~\cite{ref:antimo}.
There is in addition the possibility of having unknown resonances in the \DDK\ final states. Simulations of the known resonances indicate that the efficiencies for nonresonant modes and resonant modes are significantly different. This is due to the fact that the efficiency is not uniform across the phase space and that resonant events, depending on the mass, the width and the spin of the resonance, populate differently the Dalitz plane. Ignoring this effect would introduce a bias of up to 9\% in the total branching fraction for some decay modes
In order to measure the branching fractions inclusively without any assumptions on the resonance structure of the signal, we estimate the efficiency as a function of location
in the Dalitz plane of the squared invariant masses $m^2(\Db^{(*)}D^{(*)}) \times m^2(D^{(*)}K)$ for the data. We use this efficiency at the event position in the Dalitz plane to reweight the signal contribution. To isolate the signal contribution event-per-event, we use the \texttt{sPlot} technique~\cite{ref:splot}.

The \texttt{sPlot} technique exploits the result of the \mes\ fit (yield and covariance matrix) and the PDFs of this fit to
compute an event-per-event weight for the signal category and background category. The PDF for the signal category is $\mathcal{P}_{\mathrm{S}}$. For the background category, the three PDFs for the different components of the background (cross feed, combinatorial events, and peaking background) are combined together to form one PDF:
\begin{eqnarray}
\mathcal{P}_{\mathrm{B}} &=& \frac{N_{\mathrm{CF}} \times \mathcal{P}_{\mathrm{CF}} + N_{\mathrm{CB}} \times \mathcal{P}_{\mathrm{CB}}  + N_{\mathrm{PB}} \times \mathcal{P}_{\mathrm{PB}}}{N_{\mathrm{B}}},~~
\end{eqnarray}
where $N_{\mathrm{B}}$ is the sum of the background yields, $N_{\mathrm{B}} = N_{\mathrm{CF}} + N_{\mathrm{CB}} + N_{\mathrm{PB}}$. The PDFs and the yields are the ones obtained from the results in Sec.~\ref{sec:fit}.

The \texttt{sPlot} weight for the signal category is defined as
\begin{equation}
\label{eq:splot}
 w_{\mathrm{S}}(i) = \frac{V_{\mathrm{S},\mathrm{S}} \mathcal{P}_{\mathrm{S}}(i) + V_{\mathrm{S},\mathrm{B}} \mathcal{P}_{\mathrm{B}}(i)}
                                {N_{\mathrm{S}} \mathcal{P}_{\mathrm{S}}(i) + N_{\mathrm{B}} \mathcal{P}_{\mathrm{B}}(i)}.
\end{equation}
Here, $i$ stands for the index of the event, with $\mathcal{P}_j(i)$ $(j=\mathrm{S},\mathrm{B})$ corresponding to
the value of the PDF for the event $i$. The quantity $V_{\mathrm{S},j}$ $(j=\mathrm{S},\mathrm{B})$ is the covariance matrix
element between the signal yield $N_\mathrm{S}$ and the yield $N_j$.

The branching fraction for a specific \DDK\ mode is given by
\begin{equation}
 \label{eq:splotBF}
 \BR = \sum_i \frac{w_{\mathrm{S}}(i)}
                   {N_{\BB} \times\langle\varepsilon \BR_{\mathrm{sub}}\rangle_i},
\end{equation}
where we define
\begin{equation}
\langle\varepsilon \BR_{\mathrm{sub}}\rangle_i = \sum_j \varepsilon_{ij} \times
\BR_{\mathrm{sub},j}.
\end{equation}
The sum on $j$ is over all the $D$ subdecays of a particular \DDK\ mode. The term $\BR_{\mathrm{sub},j}$
is the product of the secondary branching fractions of the subdecay $j$:
\begin{equation}
 \BR_{\mathrm{sub},j} = \BR_{\Db^{(*)}} \times \BR_{D^{(*)}} \times \BR_K,
\end{equation}
where $\BR_{\Db^{(*)}}, \BR_{D^{(*)}}$, and $\BR_K$
are the secondary branching fractions of the $\Db^{(*)}, D^{(*)}$ and $K$ mesons~\cite{ref:pdg} (with $\BR_{K}=1$ for $K^+$ mesons). The quantity $\varepsilon_{ij}$ is the efficiency for the subdecay $j$ at the Dalitz position of event $i$. In practice, for a specific \DDK\ mode with given $D$ subdecays (\textit{e.g.} $\Dzb \to \Kp\pim \times \Dz \to \Km\pip\piz$), the efficiency is obtained by using the specific simulated signal and dividing the reconstructed signal by the generated signal in the Dalitz plane $m^2(\Db^{(*)}D^{(*)}) \times m^2(D^{(*)}K)$, which is divided in $15 \times 15$ bins for the operation. The size of the bins is roughly $0.52\times0.77 {\mathrm{\,Ge\kern -0.1em V^2\!/}c^4}$, $0.46\times0.68 {\mathrm{\,Ge\kern -0.1em V^2\!/}c^4}$, and $0.38\times0.58 {\mathrm{\,Ge\kern -0.1em V^2\!/}c^4}$ for decay modes with no $D^*$, one $D^*$ and two $D^*$ mesons respectively, depending on the available phase space.
Neighboring bins are added together if one bin contains fewer than 10 events in the reconstructed Dalitz plane. The signal is simulated assuming a flat (phase space) distribution in this Dalitz plane.

The statistical uncertainty on the branching fraction is given by~\cite{ref:splot}
\begin{equation}
 \sigma_\BR = \sqrt{\sum_i \left( \frac{w_{\mathrm{S}}(i)}
                   {N_{\BB} \times\langle\varepsilon \BR_{\mathrm{sub}}\rangle_i} \right)^2}.
\end{equation}

\subsection{Validation}

\noindent
The analysis is validated at all stages by use of MC samples. These samples consist of a mixture of continuum events and generic $B$ decays containing the \DDK\ signals with branching fractions close to the ones measured in our previous result~\cite{ref:patrick}. As a preliminary remark, it has to be noted that the analysis technique, including the selection optimization and the procedure for the fit and for the branching fraction measurement, is first determined solely on MC simulations (``blind" analysis).

First, we show that the fit is able to find the true number of simulated signal events within a $1\sigma$ interval for the 22 modes, where $\sigma$ is the statistical uncertainty reported by the fit.

Furthermore, the \texttt{sPlot} method is tested on simulated samples. It is shown that this
technique is able to tag true MC signal events with very good performance. A feature of the \texttt{sPlot} method is that the sum of \texttt{sPlot} weights for a given category is equal to the yield of this category~\cite{ref:splot}. We determine that the sum of \texttt{sPlot} signal weights for the MC signal events is equal to the number of simulated MC signal events with a relative difference smaller than 1.5\% for the majority of the modes. We also check that the sum of \texttt{sPlot} signal weights for the MC background events is compatible with zero as expected.

Finally, we perform the measurement on MC simulations and find that the analysis is able to find the branching fractions set in the simulation within a $1\sigma$ interval for most of the 22 modes, where $\sigma$ is the total uncertainty on the branching fraction (combining in quadrature statistical and systematic uncertainties). We also test the iterative procedure by randomizing the initial branching fractions and check that the branching fractions are converging to the expected values after a few iterations.

\subsection{Measurement}

\noindent
For each event, we obtain the \texttt{sPlot} weight as well as the efficiency at its Dalitz position.
Using Eq.~(\ref{eq:splotBF}), we compute the branching fraction for each of the $B$ modes.
We present these results in Table~\ref{tab:result}. We assume equal $\Bz \Bzb$ and $B^+ B^-$ production~\cite{ref:pdg}.

\section{Systematic uncertainties}
\label{sec:systSec}

\begin{table*}[htb]
 \begin{center}
  \caption{Summary of the absolute systematic uncertainties on the branching fractions for each \DDK\ mode (in units of $10^{-4}$). The values listed in this table correspond to the systematic uncertainties associated with the signal shape (a), the cross-feed contribution (b), the peaking background (c), the combinatorial background (d), the fit bias (e), the iterative procedure (f), the limited MC statistics (g), the number of bins of the Dalitz plane (h), the particle detection efficiency (i) and finally the secondary branching fractions and the number of \BB\ pairs (j). The letters in parenthesis refer to the specific paragraph in Sec.~\ref{sec:systSec}. The last column presents the total systematic uncertainties.}
 \label{tab:syst}
\vskip 0.2cm
 \begin{tabular*}{1.0\textwidth}{@{\extracolsep{\fill}} lccccccccccr}
  \hline
  \hline
\CellTop Mode & Signal & Cross &  Peaking & Comb. & Fit &
Iter. & MC & ~Bins~ & Particle & BF + & Total \\
              & shape & feed & back. & back. & ~bias~ &
~proc.~ & stat. &             &  detection & $N_{\BB}$ &  syst.  \\
 & (a) & (b) & (c) & (d) & (e) & (f) & (g) & (h) & (i) & (j) & \\
  \hline
\multicolumn{12}{c}{\CellTop $B^0$ decays through external $W$-emission amplitudes} \\
 \CellTop
\modei & 0.2 & 0.1 & 0.5 & 0.0 & 0.0 & 0.0 & 0.2 & 0.2 & 0.5 & 0.3 & 0.9 \\
\modeiv & 0.7 & 0.1 & 2.4 & 0.0 & 0.0 & 0.1 & 1.3 & 1.1 & 1.9 & 1.1 & 3.7 \\
\modev & 0.5 & 0.0 & 0.7 & 0.0 & 0.0 & 0.0 & 0.5 & 0.5 & 1.3 & 0.8 & 1.8 \\
\modeviii & 2.0 & 0.2 & 1.6 & 0.1 & 0.0 & 0.2 & 1.8 & 3.8 & 6.4 & 2.9 & 8.6 \\
\hline
 \multicolumn{12}{c}{\CellTop $B^0$ decays through external+internal $W$-emission amplitudes} \\
 \CellTop
\modeiii & 0.1 & 0.0 & 1.0 & 0.0 & 0.2 & 0.0 & 0.1 & 0.1 & 0.2 & 0.4 & 1.2 \\
\modevii & 0.9 & 0.0 & 1.3 & 0.1 & 0.1 & 0.0 & 1.3 & 0.9 & 2.4 & 2.2 & 3.9 \\
\modex & 1.2 & 0.0 & 2.0 & 0.0 & 0.0 & 0.0 & 3.2 & 2.1 & 4.2 & 2.9 & 6.7 \\
\hline
 \multicolumn{12}{c}{\CellTop $B^0$ decays through internal $W$-emission amplitudes} \\
 \CellTop
\modeii & 0.2 & 0.0 & 0.4 & 0.1 & 0.1 & 0.0 & 0.1 & 0.2 & 0.1 & 0.1 & 0.5 \\
\modevi & 0.3 & 0.4 & 3.4 & 0.1 & 0.5 & 0.1 & 0.5 & 0.3 & 0.5 & 0.3 & 3.6 \\
\modeix & 0.9 & 1.6 & 5.2 & 0.2 & 0.8 & 1.8 & 1.1 & 2.6 & 1.7 & 0.6 & 6.7 \\
\hline
 \multicolumn{12}{c}{\CellTop $B^+$ decays through external $W$-emission amplitudes} \\
 \CellTop
\modexii & 0.4 & 0.0 & 0.9 & 0.1 & 0.0 & 0.0 & 0.3 & 0.2 & 0.5 & 0.5 & 1.3 \\
\modexviii & 0.5 & 0.3 & 1.0 & 0.1 & 0.1 & 0.0 & 0.8 & 0.3 & 1.4 & 1.0 & 2.3 \\
\modexv & 0.5 & 0.3 & 2.5 & 0.3 & 0.3 & 0.3 & 0.8 & 0.5 & 1.0 & 0.7 & 3.0 \\
\modexxi & 1.6 & 1.3 & 2.5 & 0.1 & 0.1 & 0.4 & 3.1 & 5.5 & 4.9 & 2.4 & 9.0 \\
\hline
 \multicolumn{12}{c}{\CellTop $B^+$ decays through external+internal $W$-emission amplitudes} \\
 \CellTop
\modexi & 0.3 & 0.0 & 1.0 & 0.0 & 0.0 & 0.0 & 0.2 & 0.2 & 0.6 & 0.3 & 1.2 \\
\modexvii & 1.2 & 0.2 & 1.0 & 0.0 & 0.1 & 0.3 & 0.9 & 1.4 & 3.6 & 1.6 & 4.5 \\
\modexiv & 0.4 & 0.2 & 0.5 & 0.0 & 0.1 & 0.3 & 0.6 & 0.5 & 1.3 & 0.6 & 1.7 \\
\modexx & 2.0 & 0.6 & 2.8 & 0.7 & 0.2 & 1.3 & 4.2 & 6.2 & 9.0 & 3.0 & 12.6 \\
\hline
 \multicolumn{12}{c}{\CellTop $B^+$ decays through internal $W$-emission amplitudes} \\
 \CellTop
\modexiii & 0.1 & 0.0 & 0.5 & 0.0 & 0.1 & 0.0 & 0.0 & 0.0 & 0.1 & 0.1 & 0.5 \\
\modexix & 0.1 & 0.0 & 0.4 & 0.0 & 0.0 & 0.0 & 0.2 & 0.3 & 0.3 & 0.2 & 0.6 \\
\modexvi & 0.1 & 0.0 & 0.6 & 0.1 & 0.0 & 0.0 & 0.3 & 0.1 & 0.3 & 0.2 & 0.8 \\
\modexxii & 0.2 & 0.0 & 0.5 & 0.0 & 0.1 & 0.0 & 0.6 & 0.2 & 0.7 & 0.4 & 1.2 \\
\hline
\hline
 \end{tabular*}
 \end{center}
 \end{table*}

\noindent
We consider several sources of systematic uncertainties on the branching fraction measurements. Their contributions are summarized in Table~\ref{tab:syst}.

(a) We fix the width of the signal PDF from the value obtained in the fit to the signal MC sample. To estimate the systematic uncertainties originating from this choice, we repeat the fit with the width free to float for modes with high significance, namely \modevii, \modex, \modeviii\ and \modexviii. The difference observed between the width from the data and from the  MC events is roughly equal to $0.1 \mevcc$. Using this number, we repeat the fits for all modes adding $\pm 0.1 \mevcc$ to the value of the width. The difference with the nominal branching fraction gives the systematic contribution associated to the signal shape. In addition, as mentioned in Sec.~\ref{sec:totalFits}, three \DDK\ modes with low signal statistics have their PDF mean fixed to the value obtained for the simulation. We repeat the fit using the PDF mean obtained from a mode with a large statistics (namely \modex) and take the difference in branching fraction as the systematic uncertainty. These two contributions of the systematic uncertainties are added in quadrature.

(b) The cross-feed determination introduces systematic uncertainties of which two sources are identified. First, we use an alternate function for the cross-feed PDF [using a nonparametric function rather than Eq.~(\ref{eq:CF})], which gives relative systematic uncertainties below 1\%.
Second, the cross-feed branching fractions and their uncertainties are known from the results of this analysis.
To estimate the systematic uncertainties coming from this effect, we repeat the measurement applying $\pm 1 \sigma$ of the statistical uncertainty on each cross-feed contribution to a given mode. These different contributions for each cross-feed mode are then combined quadratically.

(c) The peaking background contributions are fixed from fits to the background MC simulation, using a Gaussian PDF. In these fits, the three parameters, namely, the number of events, the mean, and the width, are correlated. We generate several sets of these parameters based on the covariance matrix of the fits and recompute the branching fractions for each of these sets. From the distribution of the branching fractions, we extract the systematic uncertainties originating from the peaking background.
Another systematic effect arises from the fact that we use the MC after having scaled it to the data luminosity (using the total number of MC events passing the selection, the number of $\BB$ pairs, and the cross-section of $e^+e^- \to q\bar{q}$, where $q=u,d,s,c$). We estimate the data-MC agreement by computing the ratio of number of events for data and simulation for \mes\ between 5.22 and 5.25 \gevcc. We rescale the peaking background events using the ratio found in the specific mode (0.9 in average) and repeat the branching fraction measurement. The difference with the nominal branching fraction gives the systematic uncertainty related to this effect. We combine these two sources of uncertainties in quadrature. Given the difficulty of estimating the peaking background, this is the dominant systematic uncertainty for most of the \DDK\ modes.

(d) The systematic uncertainty associated with the assumption of a fixed value of the end point in the Argus function is estimated by repeating the fit and letting the end point free to vary in the physical region between 5.288 and 5.292~\gevcc to account for possible variations in the beam energy measurement. The difference in the branching fraction between this fit and the nominal fit gives the systematic contribution related to the combinatorial background.

(e) We investigate the fit procedure performing a large number of test fits to MC samples obtained from the PDFs fitted to the data and look for the presence of possible bias in the number of signal events. We observe that the biases
 are in most cases smaller than 10\% of the statistical uncertainty. We do not correct these small biases but take them into account in the total systematic uncertainties.

(f) An iterative procedure is performed to compute the branching fractions. We check this procedure on the MC simulation, where the results should not depend on the procedure used. The difference between the iterative and noniterative methods is small but non-negligible in some cases. We take the relative difference as the systematic contribution on the data due to the iterative procedure.

(g) The limited Monte Carlo statistics induce an uncertainty on the computation of the signal efficiency.
We use an efficiency mapping in the Dalitz plane with $15 \times 15$ bins.
To take into account this uncertainty, we
generate several efficiency mappings, where in each bin we vary the nominal efficiency
according to the efficiency uncertainty distribution. We obtain a distribution of branching fractions that we employ to determine the systematic contribution.

(h) We extract the efficiency in the Dalitz plane from a $15 \times 15$ bin mapping. We vary the numbers of bins from $1 \times 1$ bin to $20 \times 20$ bins and recompute the branching fractions. This test is performed on MC simulation containing signal and background events, where the \DDK\ signal is purely nonresonant. In this case, the results should not depend on the number of bins since no resonant states are present. The maximal difference with the nominal branching fraction is taken as the systematic uncertainty.

(i) From the differences in the reconstruction and particle identification efficiencies for the data and MC control samples, we derive systematic uncertainties of 0.2\% per charged track, 1.7\% per soft pion from $D^*$ decays, 1.2\% per \KS, 3\% per $\pi^0$, and 1.8\% per single photon. Additionally, the systematic uncertainties for the $K^+$ identification are ranging from 2\% to 4\% (in total) depending on the mode.

(j) Finally, the uncertainties on the $D^{(*)}$ and \KS\ branching fractions~\cite{ref:pdg} are accounted for. The total systematic uncertainty also takes into account the number of $B$ mesons in the data sample, which is known with a 0.6\% uncertainty.

Table \ref{tab:syst} shows a summary of the systematic uncertainties. The uncertainties from the different contributions are added together in quadrature to give the total systematic uncertainty for a specific mode.

\section{Results}

\noindent
The final results on the data using the full \babar\ data sample can be found in Table \ref{tab:result}.
In this Table, the quantity $N^{\mathrm{SR}}_{\mathrm{CF}}$ is the number of cross-feed events in the signal region (\textit{i.e.} integrating the cross-feed PDF for $\mes > 5.27 \gevcc$) determined from the MC simulation scaled to the data luminosity using the branching fractions measured in this analysis (this includes peaking and nonpeaking cross-feed contributions). We indicate the significances (including systematic uncertainties) of the observations.
To compute these significances, we repeat the fits using no contribution from the signal. We compute the statistical significance $S_{\mathrm{stat}}$ calculating \texttt{PROB}$[2 \ln (\mathcal{L}_{\mathrm{signal}}/\mathcal{L}_0),
N_{\mathrm{dof}}]$, where $\mathcal{L}_{\mathrm{signal}}$ ($\mathcal{L}_{0}$) is the maximum of the likelihood
with (without) the signal contribution, $N_{\mathrm{dof}}$ is the number of free
parameters in the signal PDF (two here), and \texttt{PROB} is the upper tail probability
of a chi-squared distribution, converting this probability into a number of standard
deviations. We then take the systematic uncertainty into account by smearing $S_{\mathrm{stat}}$ by use of a Gaussian with a width equal to the systematic uncertainty: $ S_{\mathrm{stat+syst}} = S_{\mathrm{stat}} / \sqrt{(1 + \sigma_\mathrm{syst}^2 / \sigma_\mathrm{stat}^2)}$, where $\sigma_\mathrm{stat}$ and $\sigma_\mathrm{syst}$ are, respectively, the statistical and systematic uncertainties on the branching fraction measurement.

We check isospin invariance using the \DDK\ decays. Assuming isospin invariance in the $B$ decay, interchanging the $u$ and $d$ quarks in the Feynman diagrams of Fig.~\ref{fig:diagrams} should not modify the amplitude values. Table~\ref{tab:ratios} presents the ratios of the modes which are related by isospin symmetry. In the ratio of the branching fractions, all factors cancel except the amplitudes and the $\Bz/\Bp$ lifetimes (neglecting the small mass differences between neutral and charged states for the $B$, $D^*$, $D$, and $K$ mesons). We multiply the ratios of the neutral to charged branching fractions, $r$, by the ratio of the charged to neutral $B$ meson lifetimes, $\tau_{B^+}/\tau_{B^0}=1.071 \pm 0.009$~\cite{ref:pdg}. The uncertainty on these values reported in Table~\ref{tab:ratios} combines the statistical and systematic uncertainties of our measurement, as well as the uncertainty on the lifetime ratio. The values of $r\times \tau_{B^+}/\tau_{B^0}$ should be equal to unity if isospin invariance is verified. Although some values are compatible with this equality, for some others we observe discrepancies up to $2.6\sigma$ (where $\sigma$ is the 68\% standard deviation). This result is obtained assuming equal production of $\Bz$ and $B^+$ mesons.

\begin{table*}[htb]
 \begin{center}
  \caption{Ratios of neutral to charged branching fractions. The second column shows the ratio $r$, where the first uncertainty is statistical and the second is systematic. The third column shows this value multiplied by the ratio of the charged to neutral $B$ meson lifetimes, where the error includes all uncertainties.}
 \label{tab:ratios}
\vskip 0.2cm
 \begin{tabular*}{0.8\textwidth}{@{\extracolsep{\fill}} lrr}
  \hline
  \hline
 \CellTop
  Mode & $r$~~~~~~~~~~~ & $r \times \tau_{B^+}/\tau_{B^0}$ \\
  \hline
  \CellTop
$\BR(\modei)/\BR(\modexii)$ & 0.69 $\pm$ 0.09 $\pm$ 0.08 & 0.74 $\pm$ 0.13 \\
\CellTopThree
$\BR(\modeiv)/\BR(\modexviii)$ & 0.91 $\pm$ 0.09 $\pm$ 0.11 & 0.97 $\pm$ 0.15 \\
\CellTopThree
$\BR(\modev)/\BR(\modexv)$ & 1.20 $\pm$ 0.23 $\pm$ 0.20 & 1.28 $\pm$ 0.32 \\
\CellTopThree
$\BR(\modeviii)/\BR(\modexxi)$ & 1.16 $\pm$ 0.11 $\pm$ 0.15 & 1.24 $\pm$ 0.20 \\
\CellTopThree
\vspace{1mm}
$\BR(\modeiii)/\BR(\modexi)$ & 0.57 $\pm$ 0.10 $\pm$ 0.10 & 0.61 $\pm$ 0.15 \\
$\displaystyle \frac{\BR(\modevii)}{\BR(\modexvii)+\BR(\modexiv)}$ & 0.75 $\pm$ 0.05 $\pm$
0.07 & 0.80 $\pm$ 0.09 \\
\CellTopFour
$\BR(\modex)/\BR(\modexx)$ & 0.74 $\pm$ 0.05 $\pm$ 0.10 & 0.79 $\pm$ 0.12 \\
\CellTopThree
\vspace{1mm}
$\BR(\modeii)/\BR(\modexiii)$ & 1.20 $\pm$ 0.53 $\pm$ 0.36 & 1.28 $\pm$ 0.69 \\
$\displaystyle \frac{\BR(\modevi)}{\BR(\modexix)+\BR(\modexvi)}$ & 0.88 $\pm$ 0.27 $\pm$ 0.
31 & 0.94 $\pm$ 0.44 \\
\CellTopFour
$\BR(\modeix)/\BR(\modexxii)$ & 1.81 $\pm$ 0.45 $\pm$ 0.53 & 1.94 $\pm$ 0.75 \\
\hline
\hline
 \end{tabular*}
 \end{center}
 \end{table*}

\section{Conclusion}

\noindent
We have analyzed $471 \times 10^6$  pairs of $B$ mesons produced in the \babar\ experiment, and
studied the exclusive decays of $\Bz/\Bbar^0$, \Bpm\ to \DDKpm\ and $\Bz/\Bbar^0$, \Bpm\
to $\DDK^0$. We measure the branching fractions for the 22 modes (see Table \ref{tab:result}).
Some of the modes have been observed for the first time here:  \modexiv\ ($8.3\sigma$), \modexxii\ ($7.4\sigma$), \modexix\ ($6.7\sigma$), \modexii\ ($6.6 \sigma$), and \modeiii\ ($5.1\sigma$). In addition, we show evidence for the mode \modexv\ ($3.3\sigma$) for the first time.
We also report the observation of some of the color-suppressed modes, namely \modexxii\ ($7.4\sigma$), \modexix\ ($6.7\sigma$), and \modexvi\ ($5.1\sigma$). The other color-suppressed modes are seen with a lower significance: \modexiii\ ($2.8\sigma$), \modeii\ ($2.3\sigma$), \modeix\ ($2.2\sigma$), and \modevi\ ($2.2\sigma$).

Summing the 10 neutral modes and the 12 charged modes, we measure that \DDK\ events represent $(3.68 \pm 0.10 \pm 0.24)\%$ of the $B^0$ decays and $(4.05 \pm 0.11 \pm 0.28)\%$ of the $B^+$ decays, where the first uncertainties are statistical and the second systematic, taking into account the correlations amongst the systematic uncertainties. These decays do not saturate the wrong-sign $D$ production and account roughly for one third of this production. This result implies that probably decays of the type $B \to \DDK (n \pi)$ (with $n \geq 1$) have a non-negligible contribution to the $\b \to \c \cbar \s$ transition (for example through the decays $\B \to \Dbar^{(*)} D^{(*)}  K^*$ or $\B \to \Dbar^{(*)} D^{**}  K$ where $D^{**}$ is an excited $D$ meson other than $D^{*+}$ and $D^{*0}$).

The results obtained here are found to be in satisfactory agreement with those of the previous study by \babar\ using 76 \invfb \cite{ref:patrick} and supersede these previous measurements.
Our branching fraction measurement of the mode \modex\ is found in good agreement with the values reported in Ref.~\cite{chunhui} and Ref.~\cite{ref:bellemodex}, and supersedes our previous result~\cite{chunhui}. However, our branching fraction measurement of the mode \modexi\ is in disagreement at a $2.1\sigma$ level with the Belle result~\cite{ref:belleDsJ}.

We believe that the discrepancy with Ref.~\cite{ref:belleDsJ} and the fact that the branching fractions measured here are almost systematically lower than the ones in Ref.~\cite{ref:patrick} (although most of the time compatible) are due to the fact
that in the present work we employ a more accurate parametrization of the signal \mes\ distribution in the fit and that we take into account both the cross-feed and the peaking background contributions. In Ref.~\cite{ref:patrick}, only cross feed was accounted for but the peaking background was not considered due to the lower statistics. In addition, the efficiency correction used in obtaining the branching fractions accounts for the presence of resonant intermediate states in the data.

Finally, from neutral to charged $B$ meson ratios of the branching fractions, assuming equal $\Bz \Bzb$ and $B^+ B^-$ production and taking into account the $B$ meson lifetimes, we note that some mode ratios respect the isospin invariance, while some others show discrepancies up to $2.6\sigma$.

\section*{Acknowledgments}
\noindent
We are grateful for the 
extraordinary contributions of our \pep2\ colleagues in
achieving the excellent luminosity and machine conditions
that have made this work possible.
The success of this project also relies critically on the 
expertise and dedication of the computing organizations that 
support \babar.
The collaborating institutions wish to thank 
SLAC for its support and the kind hospitality extended to them. 
This work is supported by the
US Department of Energy
and National Science Foundation, the
Natural Sciences and Engineering Research Council (Canada),
the Commissariat \`a l'Energie Atomique and
Institut National de Physique Nucl\'eaire et de Physique des Particules
(France), the
Bundesministerium f\"ur Bildung und Forschung and
Deutsche Forschungsgemeinschaft
(Germany), the
Istituto Nazionale di Fisica Nucleare (Italy),
the Foundation for Fundamental Research on Matter (The Netherlands),
the Research Council of Norway, the
Ministry of Education and Science of the Russian Federation, 
Ministerio de Ciencia e Innovaci\'on (Spain), and the
Science and Technology Facilities Council (United Kingdom).
Individuals have received support from 
the Marie-Curie IEF program (European Union), the A. P. Sloan Foundation (USA) 
and the Binational Science Foundation (USA-Israel).


\appendix

\section{Fit PDF expressions}
\label{ref:app}

We give the expressions of the PDFs introduced in Sec.~\ref{sec:fit} (along with their parameters) and used to fit the \mes\ distribution in the data.

\subsection{Signal PDF}

\noindent
The signal PDF is given by
\begin{eqnarray}
\label{eq:signal}
&\mathcal{P}_{{\mathrm{S}}}(\mes; m_{\mathrm{S}}, \sigma_{\mathrm{S}}, \alpha_{\mathrm{S}}, n_{\mathrm{S}}) = &\\
\nonumber
& \begin{cases}
\vspace{2mm}
\mathrm{exp}[-(\mes - m_{\mathrm{S}})^2/(2 \sigma_{\mathrm{S}}^2)] & \mes > m_{\mathrm{S}}-\alpha_{\mathrm{S}} \sigma_{\mathrm{S}}\\
\displaystyle \frac{(\frac{n_{\mathrm{S}}}{\alpha_{\mathrm{S}}})^{n_{\mathrm{S}}}
\mathrm{exp}(-\alpha_{\mathrm{S}}^2/2)}{((m_{\mathrm{S}}-\mes)/\sigma_{\mathrm{S}}+\frac{n_{\mathrm{S}}}{\alpha_{\mathrm{S}}}-\alpha_{\mathrm{S}})^{n_{\mathrm{S}}}} & \mes \leq m_{\mathrm{S}}-\alpha_{\mathrm{S}} \sigma_{\mathrm{S}}. \\
\end{cases} &
\end{eqnarray}
In this equation and in the following, we omit the factor that normalizes the PDF to unity.

\subsection{Cross-feed PDF}

\noindent
The cross-feed PDF for modes containing no $D^{*0}$ meson is described for the peaking part by
\begin{eqnarray}
\label{eq:CFp1}
&\mathcal{P}_{\mathrm{CF}}^{\mathrm{peaking}}(\mes; m_{\mathrm{CF}}, \sigma_{\mathrm{CF}}) =& \\
\nonumber & \mathrm{exp}[-(\mes - m_{\mathrm{CF}})^2/(2 \sigma_{\mathrm{CF}}^2)].&
\end{eqnarray}

For modes containing at least one neutral $D^{*}$ meson, the peaking component is represented by an empirical function describing an asymmetric peak:
\begin{eqnarray}
\label{eq:CFp2}
&\mathcal{P}_{\mathrm{CF}}^{\prime\ \mathrm{peaking}}(\mes; m_{\mathrm{CF}}, \sigma_{\mathrm{CF}}, t_{\mathrm{CF}}) =& \\
\nonumber & \exp\left[ \frac{\displaystyle
       -\ln^2\left(1 +
t_{\mathrm{CF}} \Lambda
         \frac{\mes-m_{\mathrm{CF}}}{\sigma_{\mathrm{CF}}}\right)}
     {\displaystyle 2\,t_{\mathrm{CF}}^2}
     - \frac{\displaystyle t_{\mathrm{CF}}^2}{\displaystyle 2}\right], &
 \end{eqnarray}
where $\Lambda = \sinh(t_{\mathrm{CF}}\sqrt{\ln4})/(t_{\mathrm{CF}}\sqrt{\ln4})$. This function approaches a Gaussian function when the parameter $t_{\mathrm{CF}}$ vanishes.

The nonpeaking part PDF is
\begin{eqnarray}
\label{eq:CFnp}
&\mathcal{P}_{\mathrm{CF}}^{\mathrm{nonpeaking}}(\mes;m_{0},\xi_{\mathrm{CF}}) =&  \\
\nonumber & \mes \sqrt{1-(\mes/m_{0})^2} \times \mathrm{exp}[-\xi_{\mathrm{CF}} (1-(\mes/m_{0})^2)].&
\end{eqnarray}

\subsection{Combinatorial background PDF}

\noindent
The combinatorial background PDF can be expressed as
\begin{eqnarray}
\label{eq:CB}
\mathcal{P}_{\mathrm{CB}}(\mes; m_{0}, \xi_{\mathrm{CB}}) &=& \mes \sqrt{1-(\mes/m_{0})^2} \\
\nonumber && \times \mathrm{exp}[-\xi_{\mathrm{CB}}(1-(\mes/m_{0})^2)].
\end{eqnarray}

\subsection{Peaking background PDF}

\noindent
The peaking background PDF is given by
\begin{eqnarray}
\label{eq:PB}
& \mathcal{P}_{\mathrm{PB}}(\mes; m_{\mathrm{PB}}, \sigma_{\mathrm{PB}}) = & \\
\nonumber & \mathrm{exp}[-(\mes - m_{\mathrm{PB}})^2/(2 \sigma_{\mathrm{PB}}^2)]. &
\end{eqnarray}

\end{document}